\documentclass[12pt,preprint]{aastex}

\shorttitle{HD 28867}
 
\begin{document}

\title{Mapping the Circumstellar Environment of T Tauri
       with Fluorescent H$_2$ Emission}
\author{Frederick M.\ Walter}
\affil{Department of Physics and Astronomy, Stony Brook University,
Stony Brook NY 11794-3800\\fwalter@astro.sunysb.edu}
\author{Gregory Herczeg}
\affil{JILA, University of Colorado,
Boulder, CO 80309-0440\\gregoryh@casa.colorado.edu}
\author{Alexander Brown}
\affil{Center for Astrophysics and Space Astronomy, University of Colorado,
       Boulder CO 80309-0389\\ab@casa.colorado.edu}
\author{David R. Ardila}
\affil{Department of Physics \& Astronomy,
             Johns Hopkins University, Baltimore MD, 21218\\
       ardila@pha.jhu.edu}
\author{G\"osta F. Gahm}
\affil{Stockholm University, AlbaNova, SE - 106 91 Stockholm, Sweden\\
      gahm@astro.su.se}
\author{Christopher M. Johns-Krull}
\affil{Department of Physics \& Astronomy, Rice University,
       Houston TX 77005-1892\\cmj@rice.edu}
\author{Jack J. Lissauer}
\affil{Space Science Division, 245-3,
       NASA/Ames Research Center, Moffett Field, CA 94035\\
       lissauer@ringside.arc.nasa.gov}
\author{Michal Simon}
\affil{Department of Physics and Astronomy, Stony Brook University,
Stony Brook NY 11794-3800\\msimon@astro.sunysb.edu}
\author{Jeff A. Valenti}
\affil{Space Telescope Science Institute, Baltimore MD 21218\\
       valenti@stsci.edu}

\begin{abstract}

We have obtained three long-slit, far UV spectra of the pre-main sequence
system T~Tauri.
These HST/STIS spectra show a strong and variable on-source spectrum
composed of both fluoresced H$_2$ and stellar chromospheric lines. 
Extended H$_2$
emission is seen up to 10\arcsec\ from the T~Tau system.
The on-source and extended H$_2$ are both  pumped by \ion{H}{1} Lyman~$\alpha$.
The on-source H$_2$ is pumped by the
red wing of a broad, self-absorbed Ly~$\alpha$ line, while the progressions
seen in the extended gas are pumped from near line center. This suggests that
the extended H$_2$ is pumped locally, and not by the stellar Ly~$\alpha$ line. 
The H$_2$ to the north and west coincides with the evacuated
cavity bounded by the optical reflection nebulosity; to the south the
extended H$_2$ coincides with the HH~255 outflow from 
the embedded infrared companion T~Tau~S.
The spatial profile of the extended gas shows a prominent dip
coincident with the position of T~Tau~S.
This may be absorption by a disk associated with T~Tau~S. 
There is no evidence for absorption by
a disk surrounding T~Tau~N large enough to obscure T~Tau~S.  

\end{abstract}
 
 
\section{Introduction}

Recent imaging and spatially-resolved spectroscopic observations of
classical T~Tauri stars and their environs
show a bewildering variety of structures, from
reflection nebulosities (e.g., McCaughrean \& O'Dell 1996)
to rings (e.g., Silber et al.\ 2000)
and edge-on disks (e.g., McCaughrean et al.\ 1998).
Some of this gas and dust represents the remnants of the birth cloud;
some is in an infalling envelope; some is in the circumstellar disk
(e.g., Adams, Lada, \& Shu 1988).
These stars often have collimated bipolar outflows orthogonal to the
disks, with entrained Herbig-Haro
nebulosities, as well as larger scale molecular outflows. The spatial scales
range from a few stellar radii to thousands of astronomical units (AU). 

The interactions between the circumstellar disk, the star, and the outflows
play a central role in the late stages of star formation.
The processes that clear away 
circumstellar disks are still not well quantified.
However, our understanding of disk dispersal will certainly 
be aided by any new 
information concerning the size and structure of gas disks and how 
central clearings and gaps develop within disks.
In addition, observations of disk material in multiple star systems
are relevant for the process of planet formation; Quintana et
al.\ (2002) have shown that if planetesimals and planetary embryos
form within disks in binary star systems, these bodies can subsequently
accrete into terrestrial planets similar to those seen within our own
Solar System.
The spatial distribution of the molecular hydrogen (H$_2$)
gas around classical T~Tauri stars 
provides important information that can address these basic questions.

Spatially extended H$_2$ emission is common in the vicinity
of young low mass pre-main-sequence (PMS) stars. 
This molecular emission is an important tracer of dynamical processes 
and shocks close to the PMS star.  Most studies published to date
have concentrated on the H$_2$ emission in the near-IR.
Despite its abundance, H$_2$ is difficult to detect from the ground
because the IR vibrational lines are forbidden quadrupole transitions and
hence are weak.
In contrast, the 
permitted UV Lyman band transitions are bright and dominate the
far UV emission from many classical T Tauri stars when observed
through large apertures (Valenti, Johns-Krull, \& Linsky 2000).
H$_2$ also dominates the far UV
emission observed from low-excitation Herbig-Haro objects
(e.g., Curiel et al.\ 1995).

H$_2$ is the major constituent of any cool gas. 
The astrophysics of H$_2$ has been reviewed by Field, Sommerville, \&
Dressler (1966) and by Shull \& Beckwith (1982).
Unlike CO, it does not
readily deplete onto dust and, thus, persists in even the densest
disk environments. H$_2$ is stable enough to resist
dissociation even at temperatures of 2-3~$\times$~10$^3$ K.
H$_2$ is a homonuclear diatomic molecule; therefore, vibrational
transitions within the X$^{1}\Sigma^{+}_{g}$
ground electronic level, falling in the near-IR, are forbidden.
However, transitions between  
X$^{1}\Sigma^{+}_{g}$ and the B$^{1}\Sigma^{+}_{u}$ and 
C$^{1}\Pi_{u}$ electronic levels are permitted,
resulting, respectively, in the Lyman and Werner bands. Emission in the
Lyman bands lies shortward of 1700~\AA. 
The HST can observe many of the Lyman band lines, while the Werner bands fall
below 1200~\AA.
The upper levels of the UV lines have lifetimes of 
10$^{-8}$~s, compared with the 10$^6$~s lifetime of the IR transitions. 
%Thus, H$_2$ emission in the UV is much stronger than in the IR. 
Also the stellar photospheric continuum is much weaker at these wavelengths,
allowing far easier detection of the H$_2$ emitting gas.
 
Valenti et al.\ (2000) detected H$_2$ emission  from 13 of the 32 T~Tauri
stars observed at low (R$\sim$200) spectral resolution with the IUE.
These Lyman band  H$_2$ emission lines are fluorescently excited 
by the hydrogen Lyman~$\alpha$ line at 1215.67~\AA\ (Brown et al.\ 1981)
and the molecular gas must be heated to $\sim$ 2000 K
for the fluorescence to operate (Brown et al.\ 1981; 
Herbst et al.\ 1996). The heating is likely a result of shocks, but X-ray
heating may also contribute (e.g., Lepp \& McCray 1983; Maloney et al.\ 1996,
Tine et al.\ 1997), especially very near the star.
Ardila et al.\ (2002a) report on the
H$_2$ emission observed in 8 classical T~Tauri stars, as observed
through the 2\arcsec$\times$2\arcsec\ aperture of the HST/GHRS.
The H$_2$ emission is fluorescently pumped by the red
wing of the stellar \ion{H}{1} Ly~$\alpha$ emission line. The lines are
narrow, single-peaked, and optically thin. The lines may be blueshifted
with respect to the stars, suggesting formation in an outflowing wind
rather than in a rotating disk.
The distribution of H$_2$ excitation temperatures appears to be
non-thermal, suggesting shock excitation of the H$_2$ into the
X$^{1}\Sigma^{+}_{g}$ state.
Herczeg et al.\ (2002) determined that the on-source H$_2$ fluorescence in the
UV spectrum of TW Hya is produced in or near the surface of the circumstellar
disk.
%Unlike many classical T~Tauri stars, TW Hya is isolated from surrounding
%molecular material and, consequently, does not show extended emission.

\subsection{T Tau}
Our target was T Tauri (HD 284419), a system that has grown more and more
complex as observations improve. The optically-visible northern component
(T~Tau~N) is a classical T~Tauri star, with evidence for on-going 
accretion. The canonical distance, which we adopt, is 140~pc. This is
consistent with the Hipparcos parallax of 5.66$\pm$1.58~mas (ESA 1997).
The star is oriented close to pole-on.
The star has a 2.8~day rotation period (Herbst et al.\ 1986) and
a rotational velocity $v$~sin~$i$ of 20.1~km~s$^{-1}$
(Hartmann et al.\ 1986). Herbst et al.\ (1986) estimated an inclination
$i$ between 8 and 13$^\circ$.
Using a different estimate of the stellar radius,
Herbst, Robberto, \& Beckwith (1997)
estimated $i$=19$^\circ$ (although they fixed $i$=10$^\circ$ for their disk
modelling); Eisl\"offel \& Mundt (1998) used the observed motions
of the Herbig-Haro flow HH~155 (32\arcsec\ west of T~Tau~N)
to derive $i$=23$^\circ$.
T~Tau~N is detected at wavelengths from X-rays through the radio; the visual
extinction A$_V$ is estimated to be 1.4-1.5~mag (Cohen \& Kuhi 1979,
Kenyon \& Hartmann 1995,
Koresko, Herbst, \& Lienert 1997, White \& Ghez 2001). 
The broadband optical and near-IR fluxes of T~Tau~N have been fairly
steady over the past half-century (Beck \& Simon 2001).
T~Tau~N has an optically thick dust disk spatially
resolved at 3~mm (Akeson, Koerner,
\& Jensen 1998), with an $\sim$40~AU radius.
Hogerheijde et al.\ (1997) found a disk radius $<$70~AU at 267~Ghz (1.1 mm).

Dyck, Simon, \& Zuckerman (1982)
showed that some of the near-IR excess arises in
the southern component (T~Tau~S).  
T~Tau~S is the prototypical infrared companion (IRC), deeply embedded
(A$_V\sim$35~mag; Koresko et al.\ 1997), and 
only visible in the IR, sub-mm and radio.
T~Tau~S dominates the flux from the system in the mid- to far-IR. 
It does not appear to be extended in mm and sub-mm maps, and indeed was
not detected by Hogerheijde et al.\ (1997) or Akeson et al.\ (1998),
but the continuum variability and 
apparently non-thermal processes argue for significant ongoing accretion.
Proposed causes for the high extinction include viewing T~Tau~S through an
edge-on disk, or viewing a background T~Tau~S through the foreground
face-on disk of T~Tau~N.
Solf \& B\"ohm (1999) argue that the outflows associated with T~Tau~S are
inclined by only about 11$^\circ$ to the plane of the sky, which supports the
notion that T~Tau~S is viewed through a nearly edge-on disk.

The T~Tau~N-S binary separation is 0.70\arcsec\ at a
position angle of 180$^\circ$ (Duch\^ene, Ghez, \& McCabe 2002).
Roddier et al.\ (2000) show a change in position angle of about
0.66$^\circ$~yr$^{-1}$. It is likely that the N-S pair is a physical
binary.
Both stars drive bipolar outflows, 
leading to the complex velocity patterns seen in optical spectra,
and powerful radio emission from the inner flow regions (Skinner 
\& Brown 1994; Ray et al.\ 1997).
Akeson et al.\
(1998) speculate that the circumstellar dust disk surrounding T~Tau~N
may be tidally truncated by interactions with
T~Tau~S. Following Hogerheijde et al.\ (1997),
Solf \& Bohm (1999) present a picture where T~Tau~S is 
viewed through an extended disk or envelope surrounding T~Tau~N.

Koresko (2000) and Duch\^ene et al.\ (2002) resolved T~Tau~S into
T~Tau~Sa and T~Tau~Sb, with a separation of about 0.09\arcsec\ in 2002.
T~Tau~Sb
appears to be a ``normal'' active low mass PMS star, with A$_V>$8~mag.
It cannot be buried within the same material obscuring T~Tau~Sa. 
Johnston et al.\ (2003) claimed that T~Tau~Sb is the T~Tau~S radio source, and
that its motions relative to T~Tau~Sa are consistent with a highly
eccentric ($e$=0.7), $\approx$40 year period orbit. The orbit does not
appear to be coaligned with the jet axis. 
Loinard, Rodr\'iguez, 
\& Rodr\'iquez (2003), accounting for the orbital motion of T~Tau~S, 
suggested that T~Tau~Sb is now moving linearly, having been ejected from
the system
during the past few years.
Tidal truncation by this close companion may explain the
absence of a spatially extended disk associated with T~Tau~S. 
Recently, Furlan et al.\ (2003) have muddied the situation by asserting that
the system is at last quadruple. Their
observations show that the T~Tau~S radio source and T~Tau~Sb are not
the same, and they suggest that a close encounter between T~Tau~Sb and the
radio source has ejected the latter.

%The summed mass of T~Tau~S is about 6.5~M$_\odot$; the
%mass of the triple T~Tau system is about 8~M$_\odot$.

The WFPC2 optical images (Stapelfeldt et al.\ 1998)
show a reflection nebulosity
extending $\sim$6\arcsec\ in the N-S direction.
The scattered
light from the bright optical star prevented them from detecting reflection
from a dust disk in the optical.

T Tau is by far the most extended and complicated PMS H$_2$ source
in the near-IR (van Langevelde et al.\ 1994). 
Herbst et al.\ (1996)
conducted K and H band imaging spectroscopy with a resolution 
of 0.7\arcsec\ and detected up to 11 quadrupole H$_2$ lines 
in their brightest knots. Herbst et al.\ (1997),
using Fabry-Perot imaging, recognized the presence of ``a complex 
system of interlocking loops and arcs within 15\arcsec\ of the central 
stars'' and estimated a mass of 10$^{-6}$ M$_\odot$ for the H$_2$
surrounding T Tau.  
Brown et al.\ (1981) showed that the Lyman band H$_2$ emission lines were
spatially extended within the 10\arcsec$\times$20\arcsec\
large aperture of the
IUE. This was confirmed by Walter \& Liu (1997;
see also Valenti et al.\ 2000), using HST observations
of T~Tau through the Large (2\arcsec) and Small (0.2\arcsec) apertures on
the GHRS instrument. 

Here we report on far~UV long-slit observations
obtained with the HST/STIS in an attempt to ascertain the origin
of this extended emission in the T~Tauri system.

\section{HST Spectral Observations}

We obtained four spectra of T Tau using the Space Telescope Imaging
Spectrograph (STIS) on the Hubble Space Telescope as part of program 8157.
Target acquisitions were
made using the CCD, through the ND3 neutral density filter. The
spectroscopic apertures are centered on T Tau~N, the visible star.
This paper concentrates on analysis of the three
long-slit spectra that we obtained through the 0.2\arcsec\ slit.
The long-slit covers 26\arcsec\ at a spatial scale of
0.0244\arcsec\ per pixel (3.4 AU per pixel).
The 50\% energy radius in the spatial profile is about 2.5~pixels
(0.06\arcsec); a Gaussian
fit to a stellar point spread function (PSF) summed from
1250 to 1700~\AA\ gives $\sigma$=3.6 pixels (0.09\arcsec).
%The spatial resolution along the slit is about 3~pixels, or 0.075\arcsec.
At the nominal 140~pc distance of T Tau, the slit width is 28~AU.
We did not search for spatial or velocity structures across the slit. 
The G140L grating provides a spectral resolution of about 2000 for a point
source; our spectral resolution for any spatially extended emission is about
a factor of 5 lower.

Although it is not important for spectra at this resolution, we have
adopted a heliocentric radial
velocity of 19~km~s$^{-1}$ for T~Tau~N (Hartmann et al.\ 1986).
This is consistent with the 17.5~km~s$^{-1}$ we obtained from spectra taken
with the Nordic Optical Telescope (NOT). We find a $v~$sin~$i$=20~km~s$^{-1}$
in the NOT spectra, in good agreement with Hartmann et al.\
(1986)\footnote{We
also obtained a high dispersion optical
spectrum using the Kitt Peak National Observatory
4m~echelle on 24 January 1997, using the standard red long setup and the T2KB
CCD for a resolution of $\approx$50,000.}.

The medium resolution (R=45000) E140M echelle spectrum was obtained through the
0.2$\times$0.06\arcsec\ (28$\times$8.4~AU) aperture. This area is comparable
to that contained in a 2 spatial pixel resolution element in the long-slit
spectra. We concentrate here on the low dispersion spectra and use the
high dispersion spectrum mainly for comparison and line identifications.
Full analysis of the E140M spectrum will appear later.

The observing log is presented in Table~\ref{tbl-obs}.

Recently, Saucedo et al.\ (2003) have reported on analysis of an independent
far~UV long-slit observation of T~Tau. They used a wider 2\arcsec\ slit,
and consequently have lower spectral resolution and could not resolve the 
H$_2$ lines as well as we do. They observed at a different position angle
(30$^\circ$).

\section{Overview of the Long-Slit Spectra}

We obtained the long-slit spectra at position angles of
345, 10, and 100 degrees.
Figures~\ref{fig-im1} and \ref{fig-im2} show two of these spatially-resolved 
spectra. Figure~\ref{fig-im1}, with the slit at a 
a position angle of 345$^\circ$, shows the spatially-resolved spectrum
corresponding to the NW-SE jet seen in the near-IR H$_2$ images
(Herbst et al.\ 1996), and the 
bright extended emission in optical forbidden 
lines such as [S II] (Solf \& B\"{o}hm 1999).
This position angle corresponds to the outflow from T~Tau~S.
Bright Lyman band H$_2$ emission is visible to 2-3\arcsec\ either side 
of the star and fainter emission extends to greater distances to the south.
The spatial distribution of H$_2$ at a position angle of 10$^\circ$ (not shown)
is very similar.
In the image at a position angle of 100$^\circ$ (Fig.~\ref{fig-im2})
the H$_2$ is significantly less extended.
This position angle corresponds to the outflows from T~Tau~N.

\subsection{The On-source Spectrum\label{sec-oss}}

The on-source spectrum samples an approximately
24~AU~$\times$~28~AU region centered on T~Tau~N.
The strongest lines in the on-source spectrum (Figure~\ref{fig-sp_on})
are those of \ion{C}{4}, \ion{Si}{4}, and \ion{He}{2}, powered by the accretion
flow, the cooler \ion{O}{1} lines from the stellar wind, and
H$_2$ lines. A list of the strongest lines, with likely identifications,
is provided in Table~\ref{tbl-onsource}.
The most pronounced difference from typical chromospheric spectra is the
presence of many Lyman band H$_2$ lines photoexcited by Ly~$\alpha$
throughout the spectrum.  Identification of emission features requires
an understanding of the expected H$_2$ emission spectrum because
the H$_2$ emission lines are blended with the
transition region, chromospheric, and wind lines,
and are often blended with each other.
 
We predicted the H$_2$ lines from the on-source spectrum by calculating the
H$_2$ emission for assumed Ly~$\alpha$ profiles.
%, and a layer of H$_2$ to reprocess the Ly~$\alpha$ emission into H$_2$ emission.  
This method is similar to the method Wood,
Karovska, \& Raymond (2002) used to analyse H$_2$ fluorescence from
the wind of Mira~B, although with their high-resolution spectra they could
calculate a Ly~$\alpha$ profile with resolved H$_2$ emission lines, while
we
assume a Ly~$\alpha$ profile from which we calculate the H$_2$ emission.
We fit the calculated H$_2$ emission to the spectrum by normalizing to the
mean of the strongest H$_2$ lines in the spectrum that occur between
1430-1540~\AA.  The H$_2$
lines below 1400~\AA\ tend to be weaker than expected from pure branching
ratios calculated by Abgrall et al.\ (1993), due to a combination of
extinction and optical depth effects.  
Because the lower levels of the short wavelength transitions have
small energies, the opacity in these lower levels can be sufficient to
weaken the flux in those transitions.
We use an extinction of A$_V=0.3$ (see \S\ref{sec-ext})
and then correct for the
optical depth effects by fitting the H$_2$ lines at short wavelengths.
Given the degeneracy between these effects, the precise value
of the extinction used, even up to A$_V=1.5$, 
does not significantly change the calculated H$_2$ spectrum.

Many different Ly~$\alpha$ emission profiles can reproduce the H$_2$
spectrum.  However, we can constrain the Ly~$\alpha$ profile
given the relative flux from different fluorescent progressions.
The resulting H$_2$ spectra are robust to the different
acceptable Ly~$\alpha$ profiles.  Certain H$_2$ lines, most notably the 0-3
P(2) line at 1279.5~\AA\ and a blend of lines at 1274~\AA, are weaker than
predicted by these models, which indicates possible problems with the
calculated H$_2$ spectrum.  
However, we consider the model acceptable for our purposes here
because the calculated
H$_2$ spectrum resembles many detected features across the G140L bandpass.
 
In the CTTS, the strongest H$_2$ progressions are those pumped by the red wing
of Ly~$\alpha$ (Ardila et al.\ 2002a, Herczeg et al.\ 2002).  This is also
true of pattern of fluorescence in the on-source spectrum of T~Tau.
The strongest on-source H$_2$ emission lines
have upper levels $v^\prime=0, J^\prime=1$, pumped at
1217.205~\AA, $v^\prime=0, J^\prime=2$, pumped at 1217.643~\AA,
and $v^\prime=1, J^\prime=12$, pumped
at 1217.904~\AA.  Emission from $v^\prime=1,
J^\prime=4$, pumped at 1216.070~\AA, is detected but weak, and emission from
$v^\prime=2, J^\prime=5$, pumped by 1-2 R(6) at 1215.726~\AA, is undetected. 
We detect no lines pumped from wavelengths
shortward of the rest wavelength of Ly~$\alpha$.
The progressions that we detect are summarized in Table~\ref{tbl-onsource_h2}.

It is possible to explain the weak flux in
progressions pumped near line center of Lyman~$\alpha$ if the
H$_2$ sees a redshifted Lyman~$\alpha$ profile from
downflowing accreting material with a velocity $>$500 km s$^{-1}$.
We consider such a large velocity unlikely, since it exceeds
the free-fall velocity.
Also, the strong accretion lines unaffected by wind absorption
(\ion{S}{3}, \ion{S}{4}, \ion{C}{4}, \ion{He}{2}, and
\ion{Cl}{1}~$\lambda$1351)
are not redshifted in the high dispersion spectrum.
Self-absorption in the Lyman~$\alpha$ line is likely to explain the observed
pattern of H$_2$ emission, if the H$_2$ lies outside the wind acceleration
region.
The strong \ion{O}{1} and
\ion{C}{2} lines in the 1300-1340~\AA\ region are strongly self-absorbed by the
stellar wind, with no emission blueward of line center.
Similar absorption likely exists in the Ly-$\alpha$ line.
Unlike the case of some other T~Tauri stars, the wind velocity of T~Tau~N is
not sufficient to absorb any strong predicted H$_2$ lines, so we cannot 
determine where the on-source H$_2$ is formed relative to the stellar wind.
Our modelling of the Lyman~$\alpha$ line profile seems to suggest a need for
a modest redshift of the emission centroid in addition to self-absorption.
%However, the 0-4 R0 and 0-4 R1 H$_2$ lines at 1333.5 and 1333.8~\AA\,
%respectively, which we would expect to be strong, appear to be absorbed by the
%\ion{C}{2} wind, and so cannot be assumed to see only a completely
%self-absorbed Lyman~$\alpha$ line profile. We suggest that a combination of
%redshifted emission plus blue self-absorption accounts for the pattern of
%H$_2$ emission lines observed.
 
After identifying the likely H$_2$ lines, we can identify other strong
emission features that are most likely produced in the accretion shock or
the wind.  The strongest emission lines are the partially resolved \ion{C}{4}
doublet at
1550~\AA\ and \ion{He}{2} emission at 1640~\AA.  The emission at 1393~\AA\ and
1402~\AA, identified in cool stars
as \ion{Si}{4} emission, is a blend of \ion{Si}{4} and H$_2$ emission.
Similarly, emission at 1335~\AA\ is a blend of \ion{C}{2} and H$_2$ emission.

Emission at 1238~\AA, which could in principle be H$_2$ emission, is more
likely \ion{N}{5} emission, although the \ion{N}{5} 1243~\AA\ line is 
not present. 
Wilkinson et al.\ (2002) detected \ion{O}{6}
emission at 1032~\AA \footnote{The 1038~\AA\ \ion{O}{6} line
is absorbed by cold H$_2$ from
$v^{\prime\prime}$=0, J=0,1,2 lower levels.
This H$_2$ is much colder than the gas we see emitting here.}, 
so we expect the accretion shock to be hot enough to
excite \ion{N}{5} emission.
A possible explanation for the absence of the 1243~\AA\ line is
absorption by a complex of four excited \ion{N}{1} lines, about 2.4 eV
above the ground state. A blueshift of about 100 km~s$^{-1}$, consistent with
a stellar wind, is needed to blanket the \ion{N}{5} 1243~\AA\ line.

The CO A~$^1\Pi-X~^1\Sigma^+$ line system lies between 1540 and 1970~\AA. These
lines are photoexcited by the strong \ion{Si}{4} and \ion{C}{4} accretion
lines. We looked for these lines, but saw no evidence of any CO emission. This
is not surprising: the abundance of CO, even assuming no depletion of the CO,
is about 10$^{-4}$ that of H$_2$.

\subsection{Variability in the On-source Spectrum}

The stellar flux from T Tau varied between these observations.
We compared
the flux in the three long-slit spectra by extracting the spectrum within
3 spatial pixels of the peak pixel ($\pm$0.085\arcsec); the total area in
the extraction aperture is 0.024~square arcsec. 

\subsubsection{Lyman $\alpha$ emission}\label{sec-lya}

There is a variable emission line centered at 1220~\AA\
(Figure~\ref{fig-lya3}),
which we identify as the red wing of the stellar \ion{H}{1}~Ly~$\alpha$
emission line. The line is not spatially extended; it is presumably 
dominated by emission from the accretion shock.
From 2000 Nov~26 to Dec~1 the line strength increased
proportionally
with the other emission lines. It was absent a month later,
and was also not seen in
the E140M observation about 3 months previous.
We conclude that the line is stellar because
it is spatially coincident with the star in two observations with position
angles differing by 125$^\circ$, and the observed wavelength does not
change. Were the source of the Ly~$\alpha$ emission not coincident with
T~Tau~N, and hence not centered within the aperture, it should be present in
the other observation at a different spatial location. 

The red edge of the Ly~$\alpha$ emission line extends to about 1223\AA.
Assuming the line is symmetric and centered at the stellar rest velocity,
the FWHM of the line corresponds to a velocity of 950~km~s$^{-1}$.
Such a broad Ly~$\alpha$ emission line is not unprecedented;
HST/STIS spectra of RU~Lupi
and DF~Tau show red emission wings at about $\lambda\lambda$1218-1222~\AA.
The unusually unobscured, pole-on star TW~Hya shows a 
broad Ly~$\alpha$ emission
profile extending about 4~\AA\ both to the blue and red of line center
(Herczeg et al.\ 2002).
For comparison, the FWHM of the \ion{Mg}{2}~$k~\&~h$ lines is about
225~km~s$^{-1}$
(Ardila et al.\ 2002b), the FWHM of the \ion{Si}{4}~$\lambda$~1393\AA\ and
\ion{C}{4}~$\lambda$~1551\AA\ lines in the high dispersion spectrum is about
200~km~s$^{-1}$, and the FWHM of the H~$\alpha$ line in our KPNO
echelle spectrum is 250~km~s$^{-1}$.

The line profiles seen on 2000 Nov~26 and Dec~1 are similar, with peak flux
near 1220~\AA.
This suggests that it is the intrinsic line strength, and not the
extinction, which is changing.
A simple model of a Gaussian line with variable line strength and fixed
$\sigma$=3.3~\AA\ absorbed by n$_H$=4.0$\times$10$^{20}$~cm$^{-2}$
reproduces the observed profiles (see \S3.3).

The H$_2$ emission is photoexcited by Ly~$\alpha$ and can therefore be used
as a proxy for Ly~$\alpha$ emission.  The H$_2$ emission on December~1 is about
60\% stronger than on November~26, which in turn was about
10\% stronger than on January~5. The sense of the variations agrees with the
observed red wing flux of Ly~$\alpha$.
The relative H$_2$ emission line strengths show no
significant differences between the 3 observations.

\subsubsection{The Accretion Continuum}

Figure~\ref{fig-c12} shows the ratio of the flux on 2000 Dec~1
to those on the 2000 Nov~26 and 2001 Jan~5. The mean flux level is about
30\% higher on Dec~1. The excess flux is blue, and is not detected longward
of about 1650~\AA, where the true stellar photospheric continuum begins to
become detectable, suggesting possible excess accretion heating as the
source. This is corroborated by the behavior of the fluxes in the strong
lines. The \ion{H}{1}~Ly~$\alpha$ emission is strongest on Dec~1
(see \S\ref{sec-lya}). The fluxes in the strong  
\ion{C}{2}, \ion{Si}{4}, and \ion{C}{4} lines are enhanced by about a factor
of 2 on Dec~1, while the cool wind lines (\ion{O}{1} and the
$\lambda\lambda$1350 and 1670~\AA\ complexes) appear to be depressed on Dec~1.
This increase in emission line strengths due to enhanced accretion 
heating is a specific example of the general case that the PMS stars with the
largest accretion rates also have the strongest emission lines (e.g.,
Johns-Krull, Valenti, \& Linsky 2000).

The mean continuum flux on Nov~26 was 1.06 times that on Jan~5.
We conclude that the accretion rate was enhanced on or about Dec~1, which
increased the accretion continuum brightness by about 30\% while increasing
the accretion heated lines by a larger factor and suppressing the
emission from the cool wind. The presence of \ion{H}{1}~Ly~$\alpha$
emission on Nov~26 suggests that the accretion episode may have been
starting about then, and was over within a month.

\subsection{The Extinction}\label{sec-ext}

The canonical visual extinction A$_V$ for T~Tau~N is 1.5~mag.
However, we believe that the true extinction column is significantly smaller
than implied by this A$_V$.

Based on the spectral lines in
the $\lambda$6495~\AA\ region (from optical echelle spectra obtained using
KPNO 4m telescope), we concur with Cohen \& Kuhi (1979)
that the spectral type is K1~IV, with
an uncertainly of about 1 spectral subtype (other authors, such as 
Bertout, Basri, \& Bouvier (1988) and White \& Ghez (2001), assign a K0
spectral type).
Interpolating between the colors
for giants and dwarfs, we expect the intrinsic $V-I_c$ color of a K1
subgiant to be about 1.0 (colors are 0.93
and 1.08 for luminosity classes
V and III, respectively). The $V-I_c$ color of 1.45 gives a color excess of
0.45~mag.
The standard reddening vector (with the
ratio of total to selective absorption R$_V$=3.1) then implies
A$_V$=1.2~mag. A $\pm$1 spectral subtype uncertainty corresponds to an
uncertainty in A$_V$ of 0.2~mag; uncertainty in the exact intrinsic color
of a K1~IV star corresponds to an uncertainty in A$_V$ of up to 0.3~mag, in
the sense that a star with surface gravity and colors
closer to a giant than a dwarf has a smaller extinction. The canonical
A$_V$=1.5~mag can be reproduced by assuming dwarf colors for unreddened
photosphere.
If any of the near-IR excess spills over into the I band the extinction will
be overestimated. 

We measured the X-ray absorption N$_X$
by fitting the ROSAT PSPC spectrum of
T~Tau with a two-component thermal model. The best fit X-ray absorption is
3.0$\times$10$^{21}$~cm$^{-2}$, which corresponds to A$_V$=1.7~mag. However,
the uncertainty on the extinction is consistent with A$_V$=0 at 1$\sigma$.

We can use information from the UV spectra, where the opacity is larger and
the effects of extinction more extreme, to check the optical extinction.

We estimate the \ion{H}{1} absorption column directly
from the shape of the 
Ly~$\alpha$ profile (\S~\ref{sec-lya}). We assume a symmetric Gaussian
emission line centered at the stellar radial velocity and self-absorbed on
the blue side. The red wing (longward of 1220~\AA\ in the December~1
spectrum) constrains the line width; we find that the
Gaussian~$\sigma$=3.3~\AA\
(FWHM=3.9~\AA). The width of the central line reversal
then sets the absorption column to be
n$_H$=4$\times$10$^{20}$~cm$^{-2}$, which corresponds to
A$_V$ of about 0.2~mag, for a normal gas-to-dust ratio and R$_V$=3.1.
This extinction is significantly smaller than the canonical value.
However, if the absorbing gas is blueshifted with respect to the
Ly~$\alpha$ emission,
this technique will underestimate the absorption column.
An absorption column of n$_H$=3$\times$10$^{21}$~cm$^{-2}$ corresponding to the
optical extinction
can match the observed profile for blueshifted
absorption velocities of $\sim$1000~km~s$^{-1}$. This is unlikely, not only
because the high required velocity exceeds that of the winds of other
PMS stars, but also because there would then be significant
absorption of the \ion{Si}{3}~$\lambda$1206\AA\ line by the blueshifted gas.

The ratio of \ion{C}{3}~$\lambda$1175 to \ion{C}{3}~$\lambda$977 fluxes
provides another constraint on A$_V$, so
we examined the FUSE spectrum of T~Tau (Wilkinson et al.\ 2002).
In the high density limit (n$_e >$ 10$^{10}$~cm$^{-3}$),
which seems reasonable for an accretion shock, we expect an intrinsic
\ion{C}{3}~$\lambda$1175/\ion{C}{3}~$\lambda$977
flux ratio near 0.6 (e.g., the CHIANTI database; Dere et al.\ 1997).
We observe a flux ratio of about 1.44 (averaging the fluxes from each of the
two FUSE detectors). For R$_V$~=~3.1, this suggests
n$_H$=7$\times$10$^{20}$~cm$^{-2}$. 
However, this line ratio is not conclusive because 
the \ion{C}{3}~$\lambda$1175~\AA\ lines are contaminated by H$_2$ Werner
band emission and the \ion{C}{3}~$\lambda$977~\AA\ line may suffer absorption
from Werner band lines arising from $\nu^{\prime\prime}$=0, and
because there is likely wind absorption in the \ion{C}{3}~$\lambda$~977~\AA\
line. Allowing for a factor of two uncertainty in the true line ratio, the
allowable column for R$_V$~=~3.1 ranges from
2$-$12$\times$10$^{20}$~cm$^{-2}$. 
This is consistent with the Ly~$\alpha$ column, but not with the optical
extinction, which corresponds to n$_H$=2.2$\times$10$^{21}$~cm$^{-2}$.

We note that TW~Hya, a system with A$_V\sim$0, has very nearly the same
\ion{C}{3}~$\lambda$1175/\ion{C}{3}~$\lambda$977 ratio (1.41). On the
assumption that TW~Hya and T~Tau~N, both face-on accreting systems, have
intrinsically similar spectra and wind absorption, then the similar line
ratios suggests similar, and small, values of the ultraviolet extinction.

The discrepancy between the optical column implied by A$_V$ and that in
the ultraviolet suggests at least three possible solutions.
The first is that the 
extinction to the site of the formation of the Ly~$\alpha$ and \ion{C}{3}
lines is less than that to the photosphere.
The second is that R$_V$ may be larger than the standard value of 3.1.
The third is that the colors do not accurately reflect the
intrinsic spectral type.

While one could imagine a geometry that would lead to significantly
inhomogeneous extinction near the star, we do not consider the first
possibility further. 

R$_V$ is often larger than the standard value of 3.1 in star forming
regions, possibly due to the presence of large grains.
We used the Cardelli et al.\ (1989) parameterization of the extinction laws
to search for a value of R$_V$ that would self-consistently agree with the
optical and UV extinctions. While a wavelength of 977~\AA\ is below the
formal 1000~\AA\ limit of their parameterization, the extrapolation to 977~\AA\
is not inconsistent with FUSE reddening results (Hutchings \& Giasson 2001). 
Adjusting R$_V$ affects the relative extinctions of the \ion{C}{3} lines and
the inferred optical column, but it does not alter the Ly~$\alpha$ column,
which is a direct measurement of the \ion{H}{1} column.
We find that the optical and \ion{C}{3} absorption columns agree,
at n$_H\sim$1.3$\times$10$^{21}$~cm$^{-2}$, for R$_V\sim$5. This column
is still about 3 times that directly measured at Ly~$\alpha$.

Why does the optical color excess imply a much larger extinction?
Intrinsic photospheric colors generally pertain to ideal stars with
uniform surface temperatures.
Gullbring et al.\ (1998) show that weak-lined T~Tauri stars
have $V-I$ colors up to 0.2~mag redder than main sequence stars of the same 
spectral type. They show that this could plausibly be due to a large filling
factor (up to 70\%) of cool starspots. While their investigation is specific
to cooler (K7-M1) stars, the effect is presumably still evident in spotted K1
stars. This effect could account for up to nearly half the observed $V-I$
color excess.

Finally, the foreground extinction can be estimated from observations
of objects that are spatially proximate to T~Tau.
Cardelli \& Brugel (1988) found the extinction to
HH~255 (Burnham's Nebula) to lie between A$_V$=0.24 (R$_V\sim$5) and
A$_V$=0.34 (R$_V$=3.2). HH~255 is the outflow from T~Tau~S; this extinction 
measures a line of sight $<$10\arcsec\ from that to T~Tau~N.
Unless there
is a very large circumstellar contribution to the T~Tau~N extinction
(T~Tau~N is viewed nearly pole-on, where circumstellar extinction
should be minimized), then the extinction to T~Tau~N may be similarly small.

The small extinction to HH~255 and the UV diagnostics suggest
that the ultraviolet extinction is indeed
much less than expected for A$_V$=1.5~mag.
On these grounds we justify our use of n$_H$=6$\times$10$^{20}$~cm$^{-2}$
for modelling the H$_2$ line spectrum.

\subsection{The Off-source Spectra}\label{sec-offss}

The spatial structure of the H$_2$ emission lines provides us with a
variety of regions from which we can extract spatially-resolved spectra.
The emission strength decreases with distance from T~Tau~N. 
The strongest discrete feature lies between 1\arcsec\ and 2\arcsec\ south of
T~Tau~N, and is seen in the images at 345$^\circ$ and 10$^\circ$ position
angles. We call this the Southern Knot. The weaker emission south of that, 
from 2\arcsec\ to 9\arcsec\ south of T~Tau~N,
is the Faint Southern Extension (FSE).
Similarly, the weak emission west of T~Tau~N at position angle 100$^\circ$
is the Faint Western Extension (FWE). We
extracted spectra from these regions and compared them to each other and to
the on-source spectrum.

The spectra seen in the FSE, FWE, and the Southern Knot are all
similar to each other (see Figure~\ref{fig-sp_comp}).
The {\it only} detected H$_2$ lines are pumped by
1-2 R(6) at 1215.726~\AA\ and 1-2 P(5) at 1216.070~\AA, 12 and 96 km s$^{-1}$,
respectively, redward of the center of the Ly~$\alpha$ line. 
These progressions are the most prominent H$_2$ lines
in sunspots (Jordan et al.\ 1978)
and in low-excitation H-H objects (Curiel et al.\ 1995; 
Schwartz 1983; Schwartz, Dopita, \& Cohen 1985).
The flux ratio of the two progressions are
similar for each of the different regions studied.  The H$_2$ lines below
1300~\AA\ are weaker, relative to lines above 1400~\AA, in the Southern
Knot than in the other regions, either due to higher extinction towards
the knot or a larger optical depth of H$_2$.

There appears to be continuum emission between the H$_2$ lines. We have
neither the spectral resolution not the S/N to determine whether this is a
\ion{H}{1} 2-photon continuum, or an H$_2$ continuum. It is most likely not a
reflection of T~Tau~N, in which case the strong transition region
and wind lines should dominate. 

This H$_2$ spectrum contrasts strongly with that seen on-source
(\S\ref{sec-oss}).
The off-source H$_2$ cannot be pumped by the same source that
pumps the on-source H$_2$; the off-source H$_2$ sees a narrow unabsorbed
Ly~$\alpha$ profile. This conclusion differs from that of Saucedo et al. 
(2003), who conclude that the FWE emission and the emission to the north
of T~Tau~N are pumped by stellar Ly~$\alpha$ from T~Tau~N.

The exact spatial regions in the FSE and the Southern Knot sampled in the two
N-S observations are not identical due to the 25$^\circ$ difference in the slit
position angle.
The H$_2$ emission detected in the FSE has the same strength in both
observations, but the continuum is 20\% stronger on December~1.  
The continuum in the Southern Knot has the same strength in both observations,
but the H$_2$ emission was about 20\% brighter on December~1 than on
January~5. 

%Because the continuum is constant, while the stellar continuum and emission
%line flux weakened significantly from December~1 to January~5,
%the continuum in the knot is most likely not reflected
%light from the star or H$_2$ emission photoexcited by stellar radiation.

\section{The Spatial Profile of the Emission}

The long-slit spectra encompass the geocoronal \ion{H}{1} Ly~$\alpha$ and 
\ion{O}{1}~$\lambda$1304~\AA\ lines. It is evident in Figures~\ref{fig-im1}
and~\ref{fig-im2} that they extend the full length of the slit.
Aside from a systematic 5\% decrease in the mean
flux level from the bottom to the top of the PA=10$^\circ$ image, there are
no significant spatial variations in the flux of the geocoronal lines.
The distribution of 
flux values is consistent with a Gaussian with $\sigma$=0.04 of the mean. 
We take this to indicate that the MAMA detector is flat-fielded to about
4\%, and that any spatial variations greater than this are unlikely to be
instrumental.

The spatial profile of the extended emission about T~Tau is highly structured.
In order to examine the spatial structure on the smaller scales of interest,
the sizes of the disk and the separation between then components, it is
necessary to separate out the stellar point spread profile.  
For a reference point spread profile we use a STIS spectrum
of the UV-bright polar AR~UMa (observation o53y01020),
observed with the identical instrumental setup.
We aligned the spatial profiles by centroiding on the seven central
pixels.
The profiles are normalized such that they have the same
median background level far from the source, and a maximum of one.

We examined the PSF for three spectral regions:
the prominent H$_2$ lines in the spectrum of the Southern Knot
(Table~\ref{tbl-sknot}),
the prominent stellar lines listed in Table {\ref{tbl-cl} (some of which
are contaminated with H$_2$ emission),
and some regions between the lines that we call continuum (Table~\ref{tbl-cl}),
but which may have a large contribution of weak line flux.
In an attempt to clearly separate the components, 
we have avoided
the $\lambda$1547~\AA\ line, the strongest in the stellar spectrum, since it
is a combination of presumably stellar \ion{C}{4} (seen on-source)
and the H$_2$ R(3) 1$\rightarrow$8 H$_2$ line, which is seen off-source.
Similarly we have avoided the lines at $\lambda\lambda$1239~\AA, 1465~\AA,
1505~\AA, and 1602~\AA, which have both stellar atmospheric and extended
H$_2$ lines at similar wavelengths.

In the two observations with the generally north-south orientation,
the bright southern knot at $-$1.3\arcsec\ dominates the extended emission,
separated from the star by a prominent dip at $-$0.7\arcsec\
(Figure~\ref{fig-lsf1}).
This dip coincides with the location of T~Tau~S.
The enhanced H$_2$ flux extends from about $-$1.7\arcsec\ to about
+0.6\arcsec. The extended emission falls off to about
2.5\arcsec\ north, while to the south the emission plateaus at a lower level
between $-$5\arcsec\ and $-$2.5\arcsec,
and reappears at a similar level between $-$9\arcsec\ and $-$7\arcsec.

We examined the inner 0.2\arcsec\ of the spatial profiles for evidence of
spatial extent. This region is
shown in Figure~\ref{fig-innerpsf}. For the H$_2$, an appropriate reference
profile is the point source profile superposed on an enhanced background, on 
the assumption that the extended emission is continuous across the position of
the star.
Both the H$_2$ and stellar line profiles
appear to be slightly extended with respect to the reference profile.
Deconvolution of the reference profiles yields H$_2$ source widths ranging from
about 1 pixel (2.8~AU) at about 80\% of the peak
flux to 8 pixels (28~AU) at 10\% of the peak flux.
This may be evidence for a brightening of the background close to the
star, perhaps associated with the disk around T~Tau~N.

The stellar profile is always less extended than the H$_2$ profile, but shows
the same overall shape. We ascribe this to weak H$_2$ 
emission lines contaminating the stellar profile.

Figure~\ref{fig-lsf2} shows the east-west profile. Weak emission is
clearly seen to the west in H$_2$, extending out about 12\arcsec.
Any spatial extent to the east is small. The extent seen in the inner
0.15\arcsec\ is similar to that seen in the north-south profile.
There is no discernable extent to the stellar line profile.

%Much of the extent may
%be attributable to extended weak H$_2$ emission or continuum flux, or to
%scattering of the strong stellar lines from dust.
%There is a significant dip in the extended stellar flux near the location
%of the H$_2$ dip, but at a slightly greater
%distance from the star. The profile in the dip is dominated by the scattered
%light from the star and from the southern knot. The relatively fainter
%%contribution of the scattered light from the southern knot in the case of the
%stellar lines allows the minumum to shift away from the star.
%Within the noise, the continuum profile is identical to the stellar line 
%profile.

\subsection{Modelling of the Spatial Profile}

Our three spatial cuts are clearly not sufficient to construct a detailed
map of the spatial distribution, but they let us examine
some very simple models for the origin of the fluorescent off-source
H$_2$ emission. The H$_2$ could lie in a
diffuse three-dimensional volume surrounding the T Tau system, it could lie
on the surface of a circumstellar disk, or, in the absence of a large
optically thick disk around T~Tau~N, could lie
on the surface of a nearby background molecular cloud.

We exclude the hypothesis that the emission arises on the surface of a
large gaseous disk surrounding T~Tau~N, because the stellar rotational
orientation suggests that any disk should be nearly face-on. If so, the
extents seen in the N-S cuts should be mirrored in the E-W cuts.

The WFPC2 image (Stapelfeldt et al.\ 1998)
appears to show optical reflection from the interior of a cavity
evacuated by the wind of T~Tau~N. Our data are broadly
consistent with this model. Since the spatially distributed H$_2$ lines
are pumped from the center of a narrow
Ly~$\alpha$ line, we seem to need a spatially distributed pumping
source. We should not assume that there is a single central pumping source,
unlike in the case of the optical image where the reflection nebulosity is
illuminated by the central star.
The H$_2$ spatial intensity distribution is a convolution
of the distribution of shocked gas with the H$_2$ gas density, modified
by foreground extinction.

We generated a simple phenomenological
model that reproduces the gross characteristics of the
spatial profile within 2\arcsec\ of T~Tau~N (Figure~\ref{fig-bmod_2}).
This model assumes a linear variation of the H$_2$ surface flux from the
southern knot to the plateau to the north. To that we add a point source
at the position of T~Tau~N and
an absorption dip centered near the position of T~Tau~S. This is all
then convolved with the point spread function.

The absorption dip is
broad, with edges sharper than a simple Gaussian function. A rectangle 
with unresolved sides and a flat bottom, after convolution
provides a satisfactory representation of the shape of the dip.
The best-fit half width is $\pm$0.32\arcsec. The flux at the bottom of the dip,
prior to convolution with the point source profile,
is about 15\% of the interpolated flat model.
We attempted to model the absorption using
softer edges, such as a Gaussian profile, but these failed to reproduce the
observed intensity profile. The requirement for a sharp-edged absorption
is robust so long as the extended emission is
reasonably smooth on spatial scales of an arcsecond or more, as appears to be
the case. This requires that absorbing medium be foreground to 85\%
of the H$_2$ emission. 

The dips visible in the two north-south orientations are similar but not
identical.
At the 345$^\circ$ position angle, we find that the absorption is centered
about $-$0.62\arcsec, while at the 10$^\circ$ position angle the dip is
centered at $-$0.74\arcsec. The uncertainties on the positions are hard
to estimate, since the central position depends strongly on the gradient of
the emission in this region. However, as is visible in Figure~\ref{fig_dips},
the 0.1\arcsec difference in offset is significant. The dip is also broader,
by about 0.1\arcsec, in the 10$^\circ$ position angle observation.

At a distance of 0.7\arcsec\ from T~Tau~N, the two north-south orientations
are separated by about 0.3\arcsec. They sample nearby, but
independent, regions on the sky. The two dips are 
consistent with absorption by a foreground extended structure. 
The north-south thickness of the absorber is 0.64\arcsec;
its length exceeds 0.5\arcsec. It is premature to deduce a shape for the 
absorber from these data, but
we suggest that this structure may be the edge-on disk of T~Tau~S
seen in absorption. 

\subsection{The Origin of the Extended H$_2$}

As noted also by Saucedo et al.\ (2003),
the distribution of H$_2$  emission revealed by our long-slit spectra
has notable similarities to structure detected in prior optical 
observations.
The WFPC2 optical images of Stapelfeldt et al.\ (1998)
show a curved reflection nebulosity illuminated
by T~Tau~N, extending $\sim$6\arcsec\ in the N-S direction. The shape of the
nebulosity is consistent with models of reflection from the
interior of a cavity evacuated
by the E-W stellar wind from T~Tau~N (Calvet et al.\ 1994).
In the Stapelfelt et al.\ model the cavity is open to the west and 
inclined to the line of sight at an angle of $\sim$ 45$^\circ$. 
In Figure~\ref{kscont} we overlay our summed H$_2$ long-slit signal 
on the Stapelfeldt et al.\ contour plot. At least along the cuts we have
sampled, the fluorescing H$_2$ appears to fill the 
volume that is foreground to the cavity wall responsible 
for the reflection nebulosity. To the north, east, and west the
extent of the H$_2$ emission is bounded by the cavity structure.

The situation is more complicated to the south, where our 
bright Southern Knot and Faint Southern Extension
emission lies well outside the structure responsible for the 
reflection nebulosity.  These emissions clearly arise from a different 
structure, known as HH~255, that 
is seen in optical long-slit spectra (Solf \& B\"ohm 1999; 
Matt \& B\"ohm 2003) and is associated with the outflow from T~Tau~S.
Structure in [\ion{S}{2}] $\lambda$6731 emission correlates very well with 
that seen in H$_2$; in particular our Southern Knot corresponds to 
a region of strong [\ion{S}{2}] that lies up-wind of a standing shock
within the jet that forms due to the jet collimation process.
The shock velocity is $\sim$90~km s$^{-1}$ and results 
in temperature high enough to produce [\ion{O}{3}] emission and destroy H$_2$.
This shock region  is seen as a steep edge in the H$_2$ 
emission located beyond $-$2.0\arcsec\ in Figure~\ref{fig-lsf1}.
It is unclear whether the faint H$_2$ emission that extends far to 
the south of this shock is related to the jet outflow or some other
diffuse gas. The [\ion{S}{2}] $\lambda$6731/$\lambda$6716 ratio suggests
that the electron density within our Faint Southern Extension is
$\sim$1$\times$10$^3$ cm$^{-3}$ (Matt \& B\"ohm 2003).  

\subsection{Implications for T~Tau S}

We speculate that the discrete dark feature at $-$0.7\arcsec\ may be the
shadow of the edge-on disk of T~Tau~S.
This is suggested primarily by spatial coincidence.
The shadow is dark; any foreground H$_2$ (or other far UV) flux contributes
no more than about 15\% of the interpolated background flux.
Note that this absorber is detected only in the far~UV, where the opacity is
relatively large;
the detection of absorbing matter in the far UV is not inconsistent with the
failure to detect extended optically thick material at mm~wavelengths.

If T~Tau~S has an extended disk, the 11$^\circ$ orientation of its rotation
axis from the plane of the sky suggests that the disk should be seen
close to edge-on, at a position angle close to perpendicular to the 
345$^\circ$ position angle of the outflow from T~Tau~S. 
We note that any elongated structure at a
position angle perpendicular to the T~Tau~S outflow will
be seen offset from T~Tau~N by 0.67 and 0.75\arcsec, respectively, in the
345$^\circ$ and 10$^\circ$ scans. This is consistent with the
observed positions of the dips in the two scans.

Any disk is likely
to be severely truncated by the presence of T~Tau~N and T~Tau~Sb.
If T~Tau~S
has a 40~AU (0.28\arcsec) radius disk like T~Tau~N, then the thickness of the
disk must be about 0.2\arcsec, a significant fraction of the radius, to 
account for the observed width of the absorption dips. We do not have the
spatial coverage to constrain the spatial extent of the absorption 
along the major axis of the possible disk. We note that
the inferred extent of the absorbing material is substantially
larger than the inferred radius of the T~Tau~Sa/Sb orbit. Clearly, an image
of this system in the light of the H$_2$ Lyman bands would greatly clarify
matters.

%At an 11$^\circ$
%inclination, the observed N-S thickness would correspond to a 1.6\arcsec\
%radius for a flat thin disk, but this is unlikely to be a flat thin disk.
%Any stable disk in this system, even given the the non-coaxial orientations,
%is likely to be truncated at a fraction of the stellar separation.

The presence of an absorber associated with T~Tau~S suggests that 
T~Tau~S is not seen through the disk of T~Tau~N because then the disk of
T~Tau~N should also cast a shadow.
While we cannot unequivocally prove that there is extended emission between
T~Tau~N and T~Tau~S, there is no comparable absorption dip to the north or
east. If the dip is due to a disk surrounding T~Tau~N, it is a most unusually
shaped disk.

If the absorption feature is a foreground structure around T~Tau~S, then
we suggest that the strong variability of T~Tau~N, visible before about 1917
(Beck \& Simon 2001), is attributable to the line of sight to T~Tau~N
passing through thick material in the equatorial plane of T~Tau~S. 
Assuming the equatorial plane is perpendicular to the outflow direction, and
assuming a 550 year circular orbit (Roddier et al.\ 2000) and no precession,
then T~Tau~N would have been in the equatorial
plane of T~Tau~S in about 1850, and perpendicular to the plane in about
1985. If there is a substantial amount of circumstellar
material associated with T~Tau~S, we would expect it to have maximally obscured
a background T~Tau~N around 1850. This is consistent with the Beck \& Simon
(2001) light curve, which shows T~Tau~N uniformly faint from 1899 through
1902 (an admittedly short interval), and then highly variable until 1917.
It is also consistent with the light curve of Lozinskii (1949), covering the
interval 1858-1941. His light curve shows long intervals between 1858 and 1870
when T~Tau~N appeared to be very faint (magnitude $\sim$13, as opposed to
$\sim$magnitude~10 after 1917).  
From 1902 to 1917 the projected perpendicular distance from T~Tau~N to the
equatorial plane of T~Tau~S increased from about 0.45\arcsec\ to 0.53\arcsec.
This is somewhat larger than our inferred projected disk thickness of
0.32\arcsec.

T~Tau~N appears not to be intrinsically highly variable: its
photographic magnitude has been essentially constant since 1926.
T~Tau~N will again lie in the equatorial plane of T~Tau~S in about the year
2175;
if the orbit is coplanar with the disk of T~Tau~N (a dubious assumption given
the absolute non-coaxiality elsewhere in the system), T~Tau~S is currently
receding from us on its orbit, and will be in the background in 2175.
We would expect that T~Tau~N will remain fairly bright for the next 325 years,
and then will once again become a highly variable star.

\section{Conclusions}

T Tauri continues to confound and perplex, but better observational data 
are yielding new insights. This is a possibly heirarchical,
but clearly a non-coaligned, system.
The rotation (outflow) axes of T~Tau~N and T~Tau~S are nearly orthogonal.
The Sa-Sb orbit appears not to lie in the equatorial plane of Sa. Understanding
the orbits of and relative distances to these stars is important in the
context of the formation of multiple stars. 

The on- and off-source H$_2$ spectra are very different.
The on-source H$_2$ is pumped by the red wing of the broad
self-absorbed Ly~$\alpha$ line produced by the accretion flow.
By contrast, the spatially extended H$_2$
lines are pumped from near the center of 
the Ly~$\alpha$ line. Since the broad stellar line is self-absorbed, the
extended warm H$_2$ must be pumped from another source.
A likely candidate is Ly~$\alpha$ generated in the same shocks that heat the
H$_2$ (e.g., Wolfire \& K\"onigl 1991). There
are at least two stars producing voluminous outflows, including HH objects,
in this system, so there is no lack of energy to drive the shocks.

The strong H$_2$ emission north and west of T~Tau appears to trace gas within 
the west-facing cavity bounded by the bright reflection
nebulosity in the WFPC2 image. The bright H$_2$ emission to the south
coincides with the pre-shock HH~255 outflow from T~Tau~S.
Future H$_2$ long-slit spectroscopy or imaging should be able to
reveal the structure of the T~Tau~S disk and jet collimation region in 
considerable detail.

If the dip in the brightness profile is related to T Tau~S, then
T~Tau~S is in front of 85\% of the fluorescing gas. It is probably not
obscured by a large disk surrounding T~Tau~N.
Rather, the irregular variability of T~Tau~N may be due to extinction
by material associated with T~Tau~S. T~Tau~N may once again become
a highly variable star in about the year 2325.

Reconstruction of the spatial distribution of the gas from the long-slit images
tantalizes one with the wealth of information suggested.
%It is clear that a far~UV image of the T~Tau system would be spectacular.
All the off-star emission shortward of 1700~\AA\ is
fluorescent H$_2$ Lyman band emission. 
A 1300-1700\AA -band image with HST/ACS
would directly reveal the location of the gas and the shocks.
It would show
the structure of the cavity, or of any other features, on spatial scales
of about 4~AU. An image would show instantly whether or
not the dark region is indeed the shadow of T~Tau~S, and give the size,
shape, and
orientation of this dark matter.
The far~UV offers a surprisingly clear view with which  
to penetrate the dusty veils of star formation.

\acknowledgments
We thank Tracy Beck for providing many insights into the T~Tau system, as well
as a critical reading of the manuscript. 
This research was funded in part by STScI grant GO-08157.01-97A
to SUNY Stony Brook, 
by STScI grant GO-08157.02-97B to the University of Colorado, 
and by the Swedish National Space Board.

This paper is based on observations made with the NASA/ESA Hubble
Space Telescope, obtained at the
Space Telescope Science Institute, which is operated by
the Association of Universities for Research in Astronomy,
Inc., under NASA contract NAS 5-26555. These
observations are associated with program 8157. 
%Support for program 8157 was provided by NASA
%through a grant from the Space Telescope Science
%Institute, which is operated by the Association of
%Universities for Research in Astronomy, Inc., under NASA contract NAS 5-26555.

%****************************************************************
\clearpage
\figcaption{The two-dimensional long-slit spectrum of T Tau, obtained on
   2000 December~1, at a position angle of 345$^\circ$. The spatial
   (Y) axis is in
   arcseconds along the slit, with positive Y corresponding to larger pixel
   numbers on the chip. The spatial
   axis corresponds approximately to the distance north of T~Tau~N.
   The intensity of the emission within 0.25\arcsec\ of T Tau~N has been
   reduced by a factor of 100 to show details near the star. The intensity
   scaling is linear.
   The dark vertical
   stripe at $\lambda$1216~\AA\ is geocoronal Ly~$\alpha$ emission;
   the notch 12\arcsec\ above T~Tau is an occulting bar in the STIS. The
   weaker geocoronal \ion{O}{1} $\lambda$1304~\AA\ emission is also evident.
   Emission in the
   strongest lines can be traced 9\arcsec\ to the south. Note the spatial
   structure of the intensity.  
\label{fig-im1}}

\figcaption{The two-dimensional long-slit spectrum of T Tau, obtained on
   2000~November~26, at a position angle of 100$^\circ$ (West on the sky
   is a positive offset along the spatial axis.) Details are as in
   Fig.~\ref{fig-im1}. Emission in the strongest lines can be seen up to
   7\arcsec\ west of the star.
\label{fig-im2}}

\figcaption{The on-source spectrum from 2001 December 1.
  This is extracted from three spatial pixels centered on T~Tau~N.
  The strongest line is a blend of \ion{C}{4} and H$_2$. Line identifications
  are given in Table~\ref{tbl-onsource}.
  \label{fig-sp_on}}

\figcaption{The region of \ion{H}{1} Ly~$\alpha$ in the stellar spectra. The
               spectra are 7 spatial pixels wide centered on T~Tau~N.
               The thick solid lines, the dotted line, and the dashed lines
               are the spectra observed on December~1,
               November~26, and January~5,
               respectively. The dash-dot trace is the mean geocoronal 
               Ly~$\alpha$ spectrum, which has been subtracted.
               The emission lines at 1206 and 1238~\AA\ are, respectively,
               \ion{Si}{3} and \ion{N}{5}. The emission lines and the
              continuum 
              are uniformly 30\% brighter on December~1. The stellar
              Ly~$\alpha$ flux is not detected on January~5.
\label{fig-lya3}}

\figcaption{Upper panel: the ratio of the stellar flux
   (within 0.085\arcsec\ of the star)
   in the December~1 observation to that on November~26.
   The continuous line is a fit to the smoothed
   data. The continuum is uniformly brighter by about 30\% shortward
   of 1600~\AA, while the strong transition region lines are enhanced 
   about a factor of 2, on December~1.
   This could be a consequence of increased accretion at that time.
  Lower panel: the ratio of the stellar flux
   on 1~December to that on January~5. The continuous
   line is a fit to the smoothed
   data. The continuum is enhanced, as in the upper plot, but there are
   more complex flux variations in the lines. Relative to the January~5
   spectrum, on December~1 \ion{H}{1}
   Ly~$\alpha$ is much more strongly enhanced relative to the
   transition region lines, while the cool wind lines ($\lambda$1300~\AA\
   \ion{O}{1} and the
   $\lambda$1350~\AA\ complex) are reduced. 
\label{fig-c12}}

\figcaption{The on-source spectrum of T Tau (bottom solid line) and the
spectrum of
the Southern Knot of T Tau (top solid line, offset by 5 and increased in
brightness by a factor of 5), from the December~1 observation.
The synthetic H$_2$ spectrum
(dotted lines) demonstrates that all of the emission lines in the Southern
Knot and many of the lines in the on-source spectrum result from H$_2$
fluorescence.  The
brightest H$_2$ lines in the on-source spectrum are not detected in the
Southern Knot.
%The on-source spectrum from 2001 December 1. The three panels
%  show the spectra extracted from 3 regions, as defined in the text. The
%  solid line, dashed line, and dotted line respectively represent
%  T Tau~N (on-source), the southern knot, and the weak extended emission
%  south of the knot. 
  \label{fig-sp_comp}}

\figcaption{Comparison of the observed spatial profiles of the 
H$_2$ lines in the R(3)/P(5) and R(6)/P(8) progressions (thick line), the
stellar lines (thin line) and the reference point source profile (AR~UMa;
dotted line).
These data are for the 345$^\circ$ position angle; the profiles at the
10$^\circ$ position angle are similar.
To the north (positive displacement) the H$_2$ emission is clearly
extended, while the
stellar lines appear to be enhanced relative to the reference profile.
To the south the Southern
Knot at -1.3\arcsec\ is prominent in both sets of lines.
There is no evidence for significant extended emission between the Southern
Knot and the star (the apparent excess in the H$_2$ emission above the
reference profile between -0.5 and -0.7\arcsec\ can be accounted for by
scattering from the Southern Knot),
as would be expected if that region is shadowed by a foreground obscuration.
\label{fig-lsf1}}

\figcaption{The spatial profiles on an expanded scale to show the behavior
close to the star. The H$_2$ profile is the thick solid line; the stellar
line profile is the thin solid line. The two reference profiles are plotted as
dotted lines. The lower reference profile is a point source without any
background; the upper reference profile is the same point source but
superposed upon a flat background fit to the H$_2$ profile at -1.4 and
+0.4\arcsec. The latter reference profile should be compared with the H$_2$
profile. The H$_2$ and stellar line profiles are broader than
the reference profiles.
\label{fig-innerpsf}}

\figcaption{As Figure~\ref{fig-lsf1}, but for the 100$^\circ$ orientation
observation. East is to the left (negative displacement). 
\label{fig-lsf2}}

\figcaption{A simple model (smooth curve between $-$2 and +3~arcsec)
for the spatial distribution of the extended H$_2$
emission in the north-south direction. The extended emission is modeled as a
linear brightness disribution.
The dip at -0.62\arcsec, near the position of T~Tau~S, has a half width of
0.32\arcsec, sharp edges, and is black at the center prior to convolution with
the instrumental resolution. This model is invalid outside the plotted range.
\label{fig-bmod_2}}

\figcaption{Comparison of the spatial profiles at position angles of
345$^\circ$ (thick line) and 10$^\circ$ (thin line). Fluxes are normalized to
the peak flux in the December~1 (345$^\circ$) observation.
The dip near -0.7\arcsec\ is displaced to greater negative positions, and is
somewhat broader, in the 10$^\circ$ cut. 
\label{fig_dips}}

\figcaption{The logarithmically-scaled total observed intensity in the
spatially resolved far~UV spectra, overlaid on the optical reflected
light contours (Stapelfeldt et al.\ 1998). To the north and the east the
far~UV flux does not extend beyond the edges of the reflected light, while
the Southern Knot coincides with the HH~255 outflow.
H$_2$ emission seems to fill the
open cavity to the west. The plus marks the position of T~Tau~S.
\label{kscont}}

%****************************************************************
\clearpage
\begin{deluxetable}{llllrr}
\tablecolumns{6}
\tablewidth{0pt}
\tablecaption{Observing Log\label{tbl-obs}}
\tablehead{
\colhead{Root} & \colhead{Date} & \colhead{Aperture}& \colhead{Grating} &
  \colhead{Exposure} & \colhead{PA} \\
               & & & & \colhead{(s)}      & \colhead{$^\circ$}}
\startdata
o5e301 & 2000 Dec 01 & 52x0.2 & G140L & 6800 & 345 \\
o5e302 & 2000 Nov 26 & 52x0.2 & G140L & 6800 & 100 \\
o5e303 & 2001 Jan 05 & 52x0.2 & G140L & 7200 &  10 \\
o5e304 & 2000 Sep 08 & 0.2x0.06 & E140M & 12080 & --  \\
\enddata
\end{deluxetable}

\clearpage
\begin{deluxetable}{llrl}
\tablecolumns{4}
\tablewidth{0pt}
\tablecaption{Strong Lines in the on-Source Spectrum\label{tbl-onsource}}
\tablehead{
\colhead{Wavelength} & \colhead{ID} & \colhead{Flux\tablenotemark{a}} &
       \colhead{ }\\
                & & \colhead{(10$^{-15}$~erg cm$^{-2}$ s$^{-1}$)}}
\startdata
1175.4 & \ion{C}{3} & 8.7 \\
1206.3 & \ion{Si}{3} & 8.2 \\
1220   & \ion{H}{1} &  & b\\
1238.5 & \ion{N}{5}, H$_2$ 2-2 R(11) & 18.6\\
1265.6 & \ion{C}{1}, H$_2$ 4-1 P(19) & 5.6 \\
1270.5 & \ion{S}{1}, H$_2$ 2-2 P(13), 2-2 R(14) & 3.0 \\
1274.4 & H$_2$ 0-3 R(0), 0-3 R(1) & 7.5 \\
1282.8 & H$_2$ 0-3 P(3) & 4.2 \\
1295.7 & \ion{S}{1} & 26.4 \\
1302.9 & \ion{O}{1} & 41.8 \\
1305.9 & \ion{S}{1}, \ion{O}{1} & 36.6\\
1309.3 & \ion{Si}{2} & 18.5\\
1324.8 & H$_2$ 2-3 R(14) & 3.0\\
1335.5 & \ion{C}{2}, H$_2$ 0-4 R(0), 0-4 R(1), 0-4 P(2) & 35.1 & c\\
1341.8 & H$_2$ 0-4 P(3) & 11.1\\
1351.5 & \ion{Cl}{1} & 3.1 \\
1355.3 & \ion{O}{1},\ion{C}{1} & 5.4 \\
1359.4 & \ion{C}{1} & 9.9 \\
1365.9 & H$_2$ 4-3 P(19) & 4.0 \\
1371.1 & H$_2$ 2-5 P(16) & 2.8 \\
1393.7 & \ion{Si}{4}, H$_2$ 0-5 R(0), 0-5 R(1) & 34.9 \\
1399   & H$_2$ 0-5 P(2) & & d\\
1401.8 & \ion{Si}{4}, H$_2$ 0-5 P(3) & 47.6 \\
1434.8 & H$_2$ 2-5 P(13) & 14.1\\
1444.9 & H$_2$ 1-6 P(5) & 5.0\\
1455.1 & H$_2$ 0-6 R(1), 0-6 R(2), 2-6 R(11) & 20.9 \\
1463.9 & H$_2$ 0-6 P(2), 0-6 P(3) & 21.1 \\
1472.3 & \ion{S}{1} & 19.2 \\
1482.5 & \ion{S}{1}, H$_2$ 0-6 R(14) & 5.1 \\
1485.6 & \ion{S}{1}, H$_2$ 0-5 P(13) & 2.8 \\
1488.7 & H$_2$ 1-7 R(3) & 7.5 \\
1504.9 & H$_2$ 1-7 P(5) & 7.7 \\
1516.4 & H$_2$ 0-7 R(0), 0-7 R(1)& 7.3 \\
1525.3 & H$_2$ 0-7  P(3), \ion{Si}{2} & 11.6 \\
1534.5 & \ion{Si}{2} & 6.0 \\
1547.8 & \ion{C}{4} & 111. \\
1550.6 & \ion{C}{4} & 44.9 \\
1555.5 & H$_2$ 2-8 R(11) & 3.2 \\
1562.2 & H$_2$ 1-8 P(5)  & 4.5 \\
1588.7 & H$_2$ 2-8 P(13) & 6.5 \\
1602.1 & H$_2$ 2-9 R(11) & 15.7\\
1632.1 & H$_2$ 2-9 P(13) & 5.5 \\
1640.4 & \ion{He}{2}  & 52.3 \\
1649.0 & \ion{Fe}{2}  &  1.5 \\
1660.0 & \ion{O}{3}] &  6.4 \\
1665.6 & \ion{O}{3}] & 14.1 \\
\enddata
\tablenotetext{a}{Fluxes observed on 1 December 2000. Fluxes of
blended lines were measured by fitting multiple Gaussians.}
\tablenotetext{b}{Combination of geocoronal and stellar Ly~$\alpha$. See
                  \S\ref{sec-lya}.}
\tablenotetext{c}{Unresolved doublet.}
\tablenotetext{d}{The line is noticeable on wing of \ion{Si}{4} line, but flux
                  cannot be measured accurately at this dispersion.}
\end{deluxetable}

\clearpage

\begin{deluxetable}{rrc}
\tablecolumns{3}
\tablewidth{0pt}
\tablecaption{H$_2$ Progressions in the on-Source Spectra\label{tbl-onsource_h2}}
\tablehead{
\colhead{Wavelength} & \colhead{ID} & Blend\tablenotemark{a}\\
       \colhead{(\AA)}   &} 
\startdata
\cutinhead{Pumped by 2-1 P(13) at 1217.904}
1271 & 2-2 P(13) & H$_2$\\
1325 & 2-3 P(13) & H$_2$\\
1434 & 2-5 P(13)\\
1556 & 2-8 R(11)\\
1589 & 2-8 P(13)\\
1602 & 2-9 R(11)\\
1632 & 2-9 P(13)\\
%1238 & 2-2 R(11)\\
%1453 & 2-6 R(11)\\
\cutinhead{Pumped by  0-2 R(1) at 1217.643}
1275 & 0-3 R(1) & H$_2$\\
1283 &  0-3 P(3) \\
1334 &  0-4 R(1) & C II, H$_2$\\
1342 & 0-4 P(3) &\\
1394 & 0-5 R(1) & Si IV, H$_2$\\
1403 & 0-5 P(3) & Si IV\\
1455 & 0-6 R(1) & H$_2$\\
1464 & 0-6 P(3) \\
1516 & 0-7 R(1) & H$_2$\\
1525 & 0-7 P(3)\\
1576 & 0-8 R(1)\\
\cutinhead{Pumped by 1-2 P(5) at 1216.070}
1431 & 1-6 R(3)\\
1446 & 1-6 P(5)\\
1490 & 1-7 R(3)\\
1505 & 1-7 P(5)\\
1562 & 1-8 P(5)\\
\cutinhead{Pumped by 0-2 R(0) at 1217.205}
1274 & 0-3 R(0) & H$_2$\\
1279 & 0-3 P(2) & H$_2$\\
1333 & 0-4 R(0) & C II, H$_2$\\
1338 & 0-4 P(2) & C II, H$_2$\\
1394 & 0-5 R(0) & Si IV, H$_2$\\
1399 & 0-5 P(2) & Si IV\\
1455 & 0-6 R(0) & H$_2$\\
1460 & 0-6 P(2)\\
1516 & 0-7 R(0) & H$_2$\\
1521 & 0-7 P(2)\\
\cutinhead{Pumped by 2-1 R(14) 1218.536}
1257 & 2-1 P(16)\\
1271 & 2-2 R(14) & H$_2$\\
1324 & 2-3 R(14) & H$_2$\\
1481 & 2-6 R(14) & S I\\
\cutinhead{Pumped by 4-0 P(19) 1217.410}
1267 & 4-1 P(19)\\
1274 & 4-2 R(17) & H$_2$\\
1372 & 4-4 R(17)\\
\cutinhead{Pumped by 0-3 R(2) 1219.089}
1407 & 0-5 P(4)\\
\cutinhead{Pumped by 2-2 R(9) 1219.089}
1573 & 2-8 R(9)\\
1592 & 2-9 R(9)\\
\cutinhead{Pumped by 2-2 P(8) 1219.154}
1579 & 2-9 R(6)\\
\enddata
\tablenotetext{a}{Other species in line blend}
\end{deluxetable}

\clearpage
\begin{deluxetable}{rr}
\tablecolumns{2}
\tablewidth{0pt}
\tablecaption{Strong Features in the Southern Knot\label{tbl-sknot}}
\tablehead{
\colhead{Wavelength} & \colhead{ID}\\
  \colhead{(\AA)}   & }
\startdata

1238.0 & P(8) 1-2\\
1258.0 & R(3) 1-3\\
1272.5 & P(5) 1-3 +  R(6) 1-3\\
1293.7 & P(8)   1-3\\
1315.2 & R(3)   1-4\\
1373.0 & R(3)   1-5\\
1387.6 & P(5)   1-5\\
1432.0 & R(3)   1-6\\
1443.0 & R(6) 1-6 + P(5) 1-6\\
1446.5 & P(5) 1-6 + R(6) 1-6\\
1467.5 & P(8)   1-6\\
1489.9 & R(3)   1-7\\
1500.5 & R(6)   1-7\\
1504.5 & P(5)   1-7\\
1524.9 & P(8)   1-7\\
1547.3 & R(3)   1-8\\
1557.0 & R(6) 1-8\\
1562.9 & P(5)   1-8\\
1579.5 & P(8)   1-8\\
1603.1 & R(3)   1-9\\
1618.3 & P(5)   1-9\\
\enddata
\end{deluxetable}

\clearpage
\begin{deluxetable}{rr}
\tablecolumns{2}
\tablewidth{0pt}
\tablecaption{Features Used for Spatial Profile Analysis\label{tbl-cl}}
\tablehead{
\colhead{Wavelength} & \colhead{ID}\\
  \colhead{(\AA)}   & }
\startdata
\cutinhead{Wavelengths of stellar lines}
1176.3 & \ion{C}{3}\\
1206.5 & \ion{Si}{3}\\
1239.2 & \ion{N}{5} + H$_2$\\
1336.2 & \ion{C}{2} + H$_2$\\
1357.7 & \ion{C}{1} + \ion{O}{1}\\
1394.2 & \ion{Si}{4} + H$_2$\\
1401.6 & \ion{Si}{4} + H$_2$\\
1455.4 & H$_2$ R(0) 0-6 \\
1472.3 & \ion{S}{1}\\
1547.3 & \ion{C}{4}  + H$_2$\\
1590.4 & H$_2$\\
1641.9 & \ion{He}{2}\\
1666.3 & \ion{C}{1} + \ion{Fe}{2} + \ion{O}{3]}  \\
\cutinhead{Continuum regions}
\multicolumn{2}{c}{1682.2 $\pm$ 10.2}\\
\multicolumn{2}{c}{1707.5 $\pm$  2.8}\\
\enddata
\end{deluxetable}

%%%%%%%%%%%%%%%%%%%%%%%%%%%%%%%%%%%%%%%%%%%%%%%%%%%%%%%%%%%%%%%%
%%%                  FIGURES
%%%%%%%%%%%%%%%%%%%%%%%%%%%%%%%%%%%%%%%%%%%%%%%%%%%%%%%%%%%%%%%%
 
\begin{figure}
\plotone{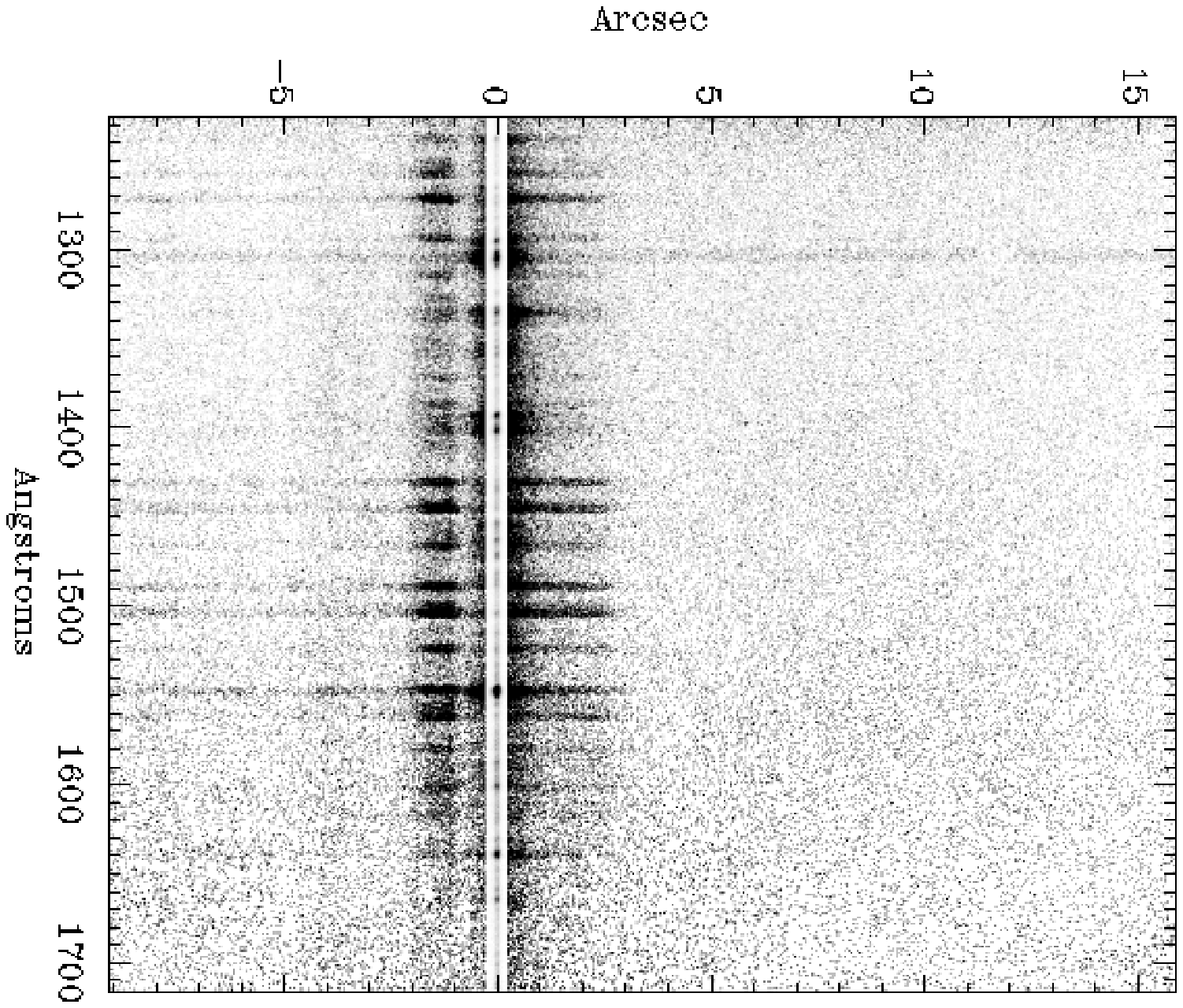}
\end{figure}

\begin{figure}
\plotone{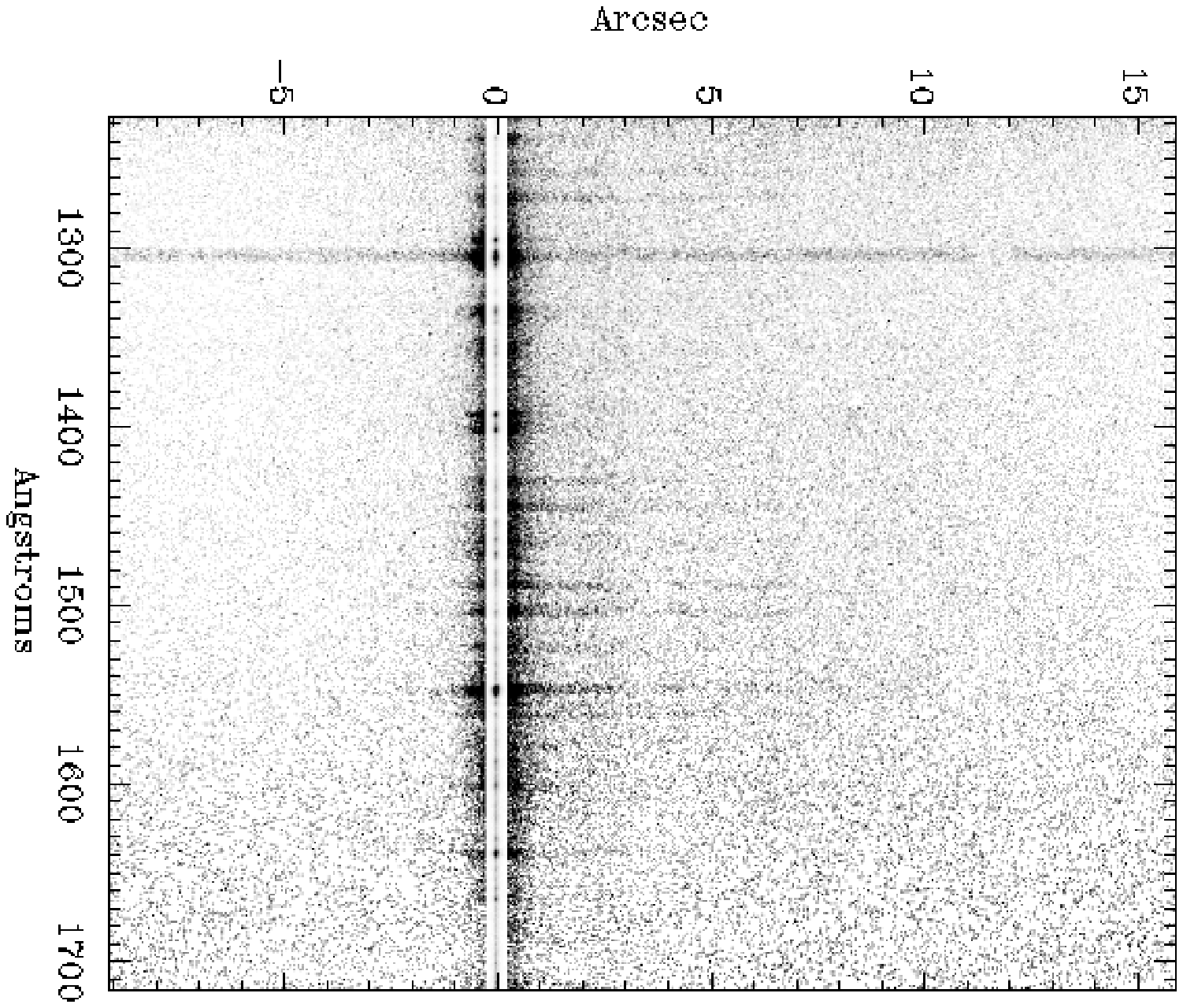}
\end{figure}

\begin{figure}
\plotone{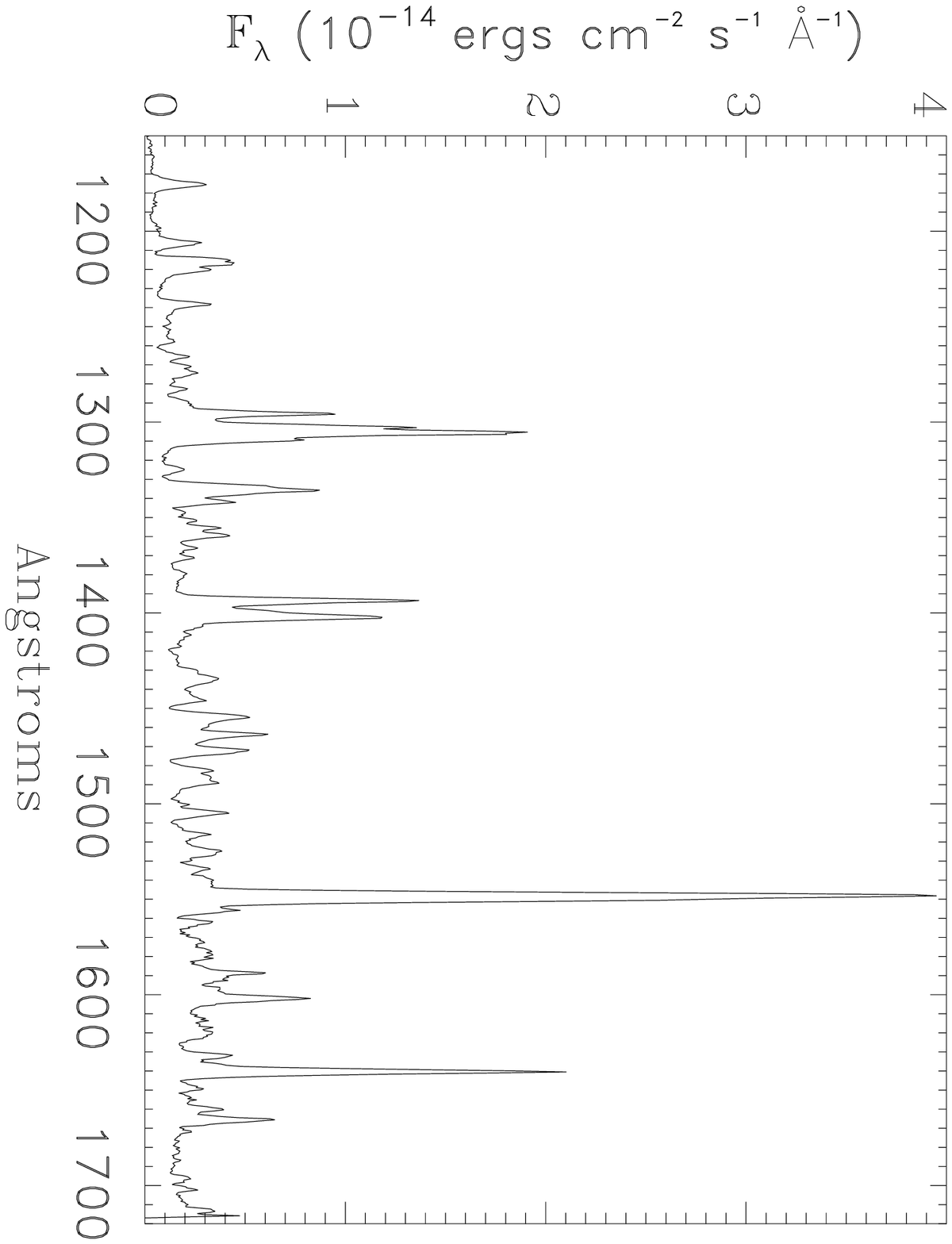}
\end{figure}

\begin{figure}
\plotone{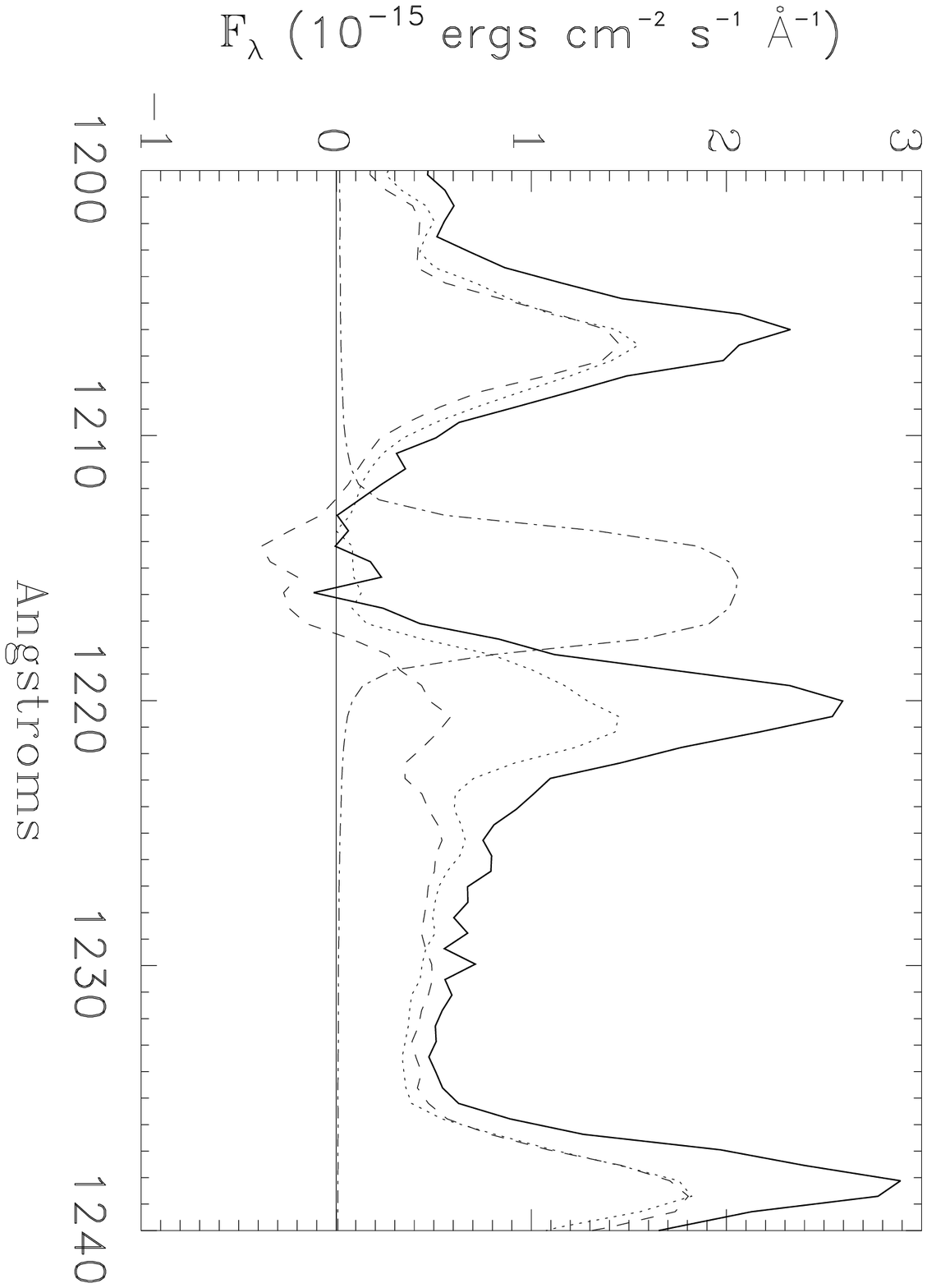}
\end{figure}

\begin{figure}
\plotone{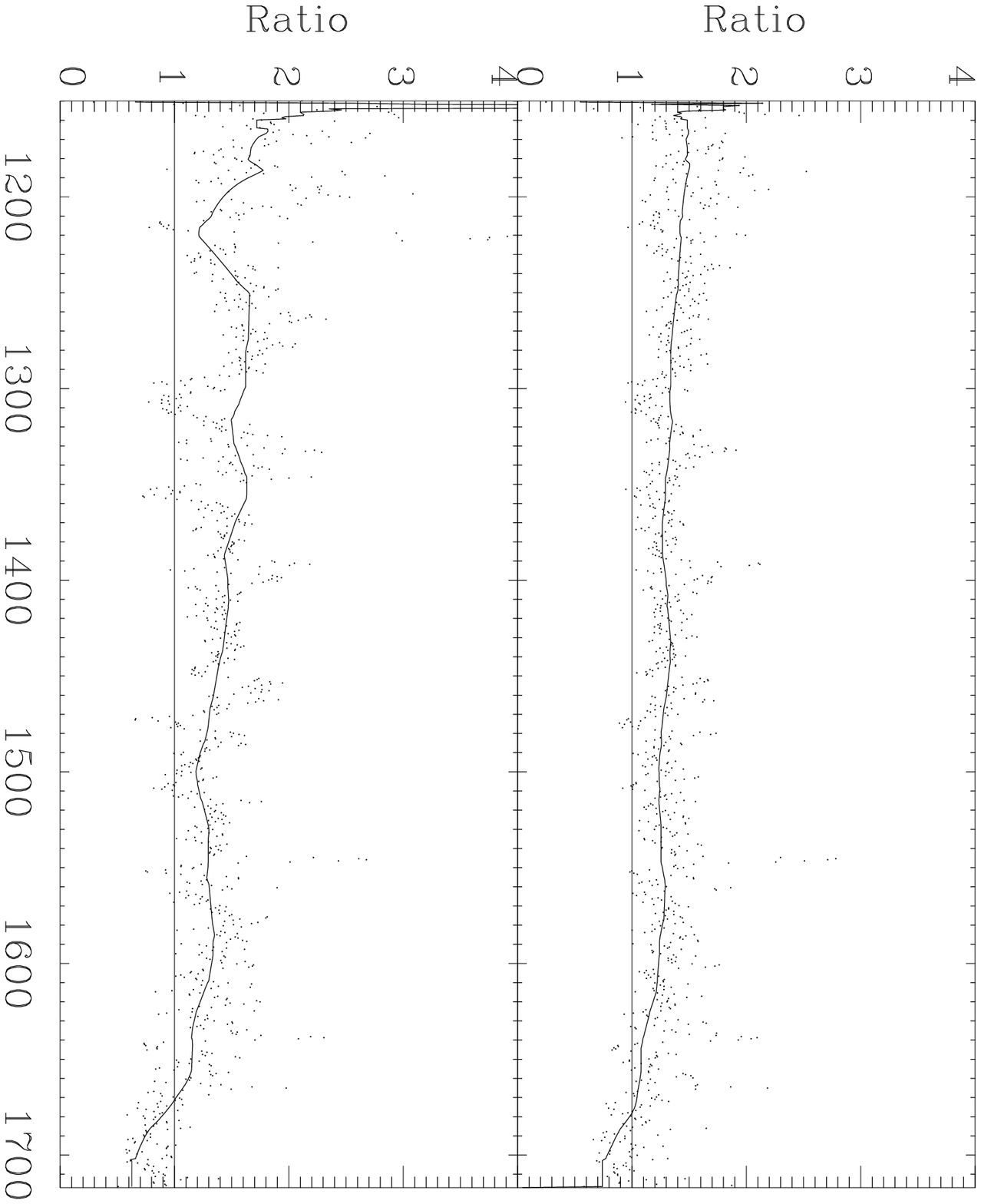}
\end{figure}

\begin{figure}
\plotone{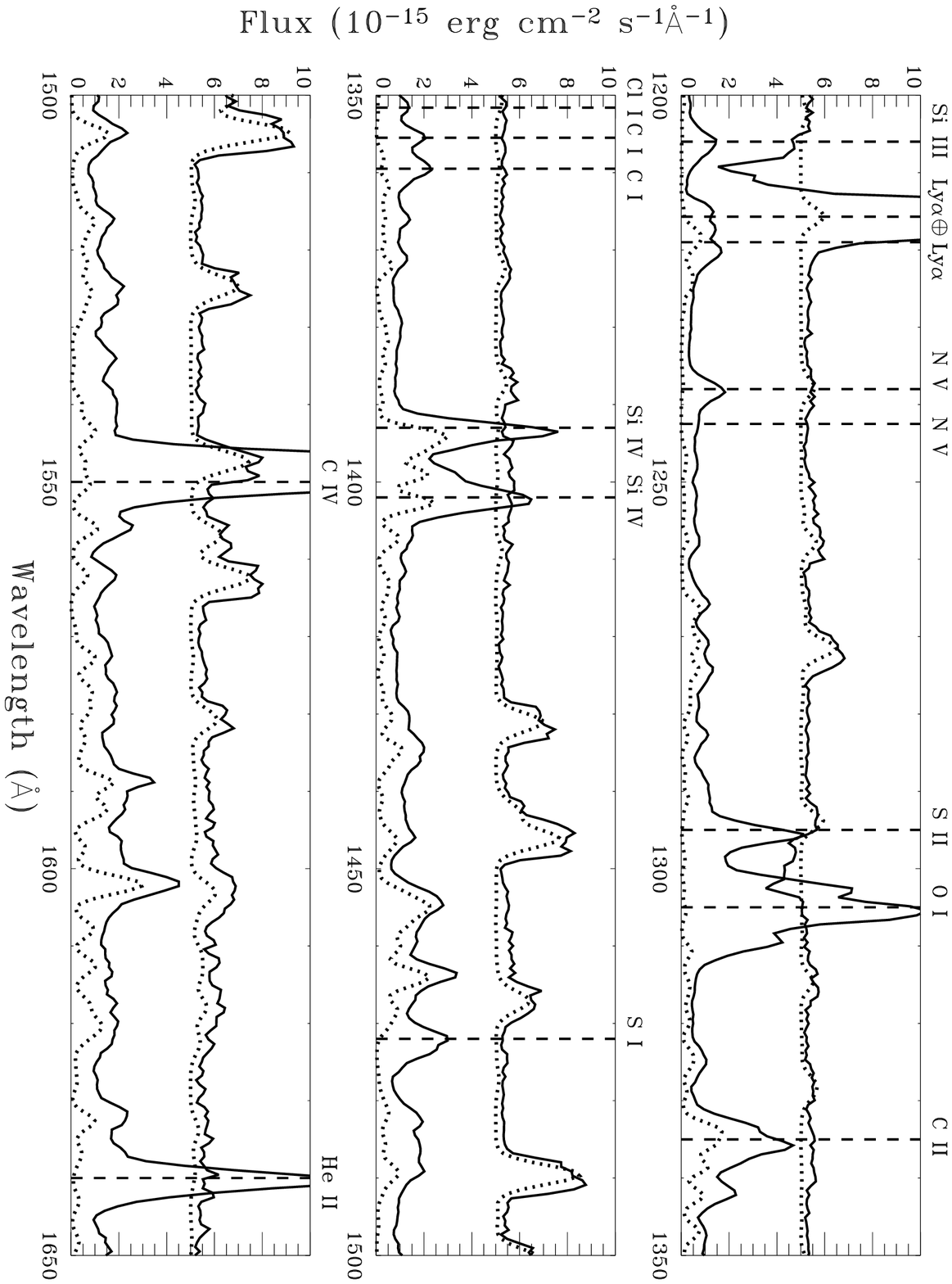}
\end{figure}

\begin{figure}
\plotone{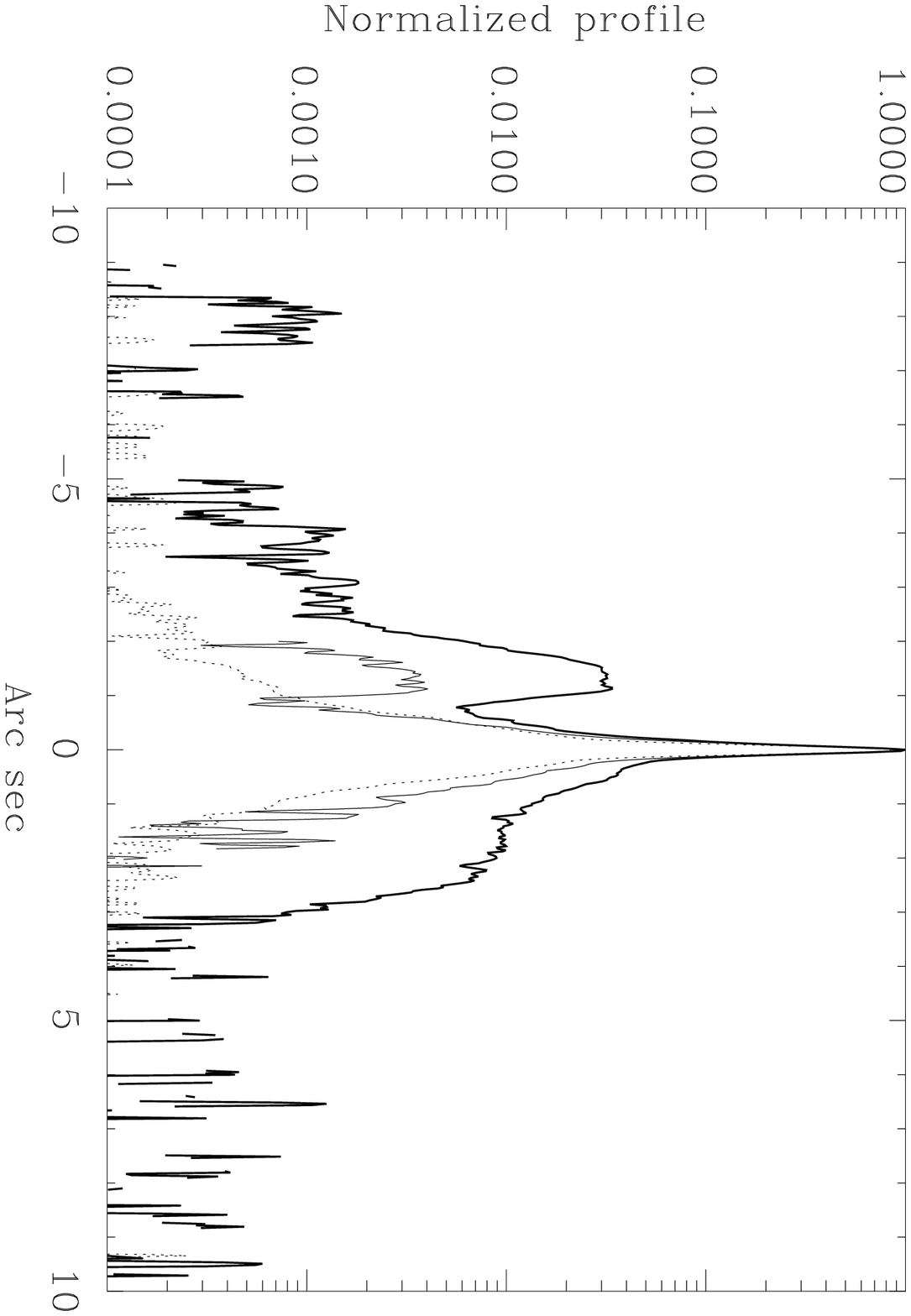}
\end{figure}

\begin{figure}
\plotone{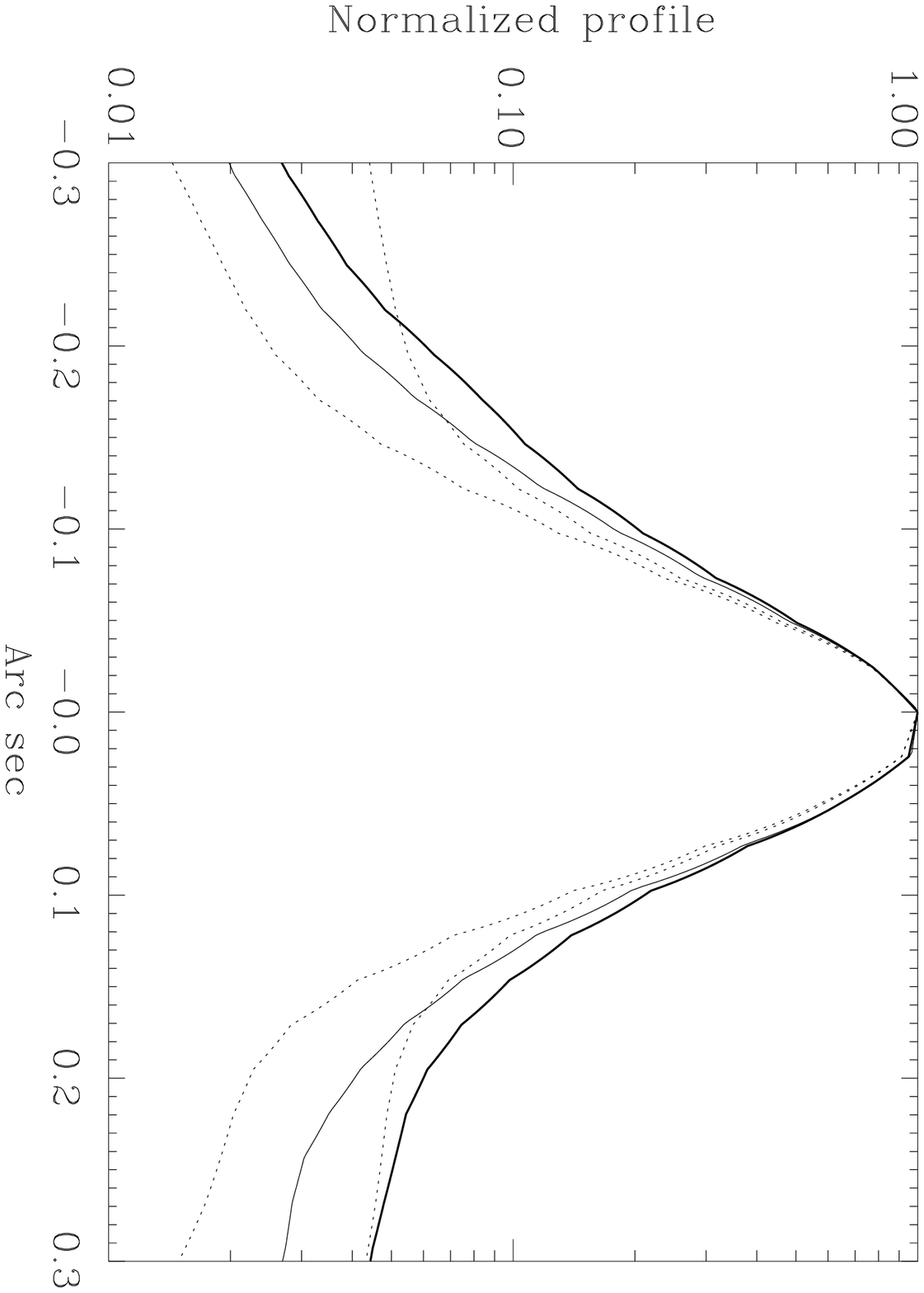}
\end{figure}

\begin{figure}
\plotone{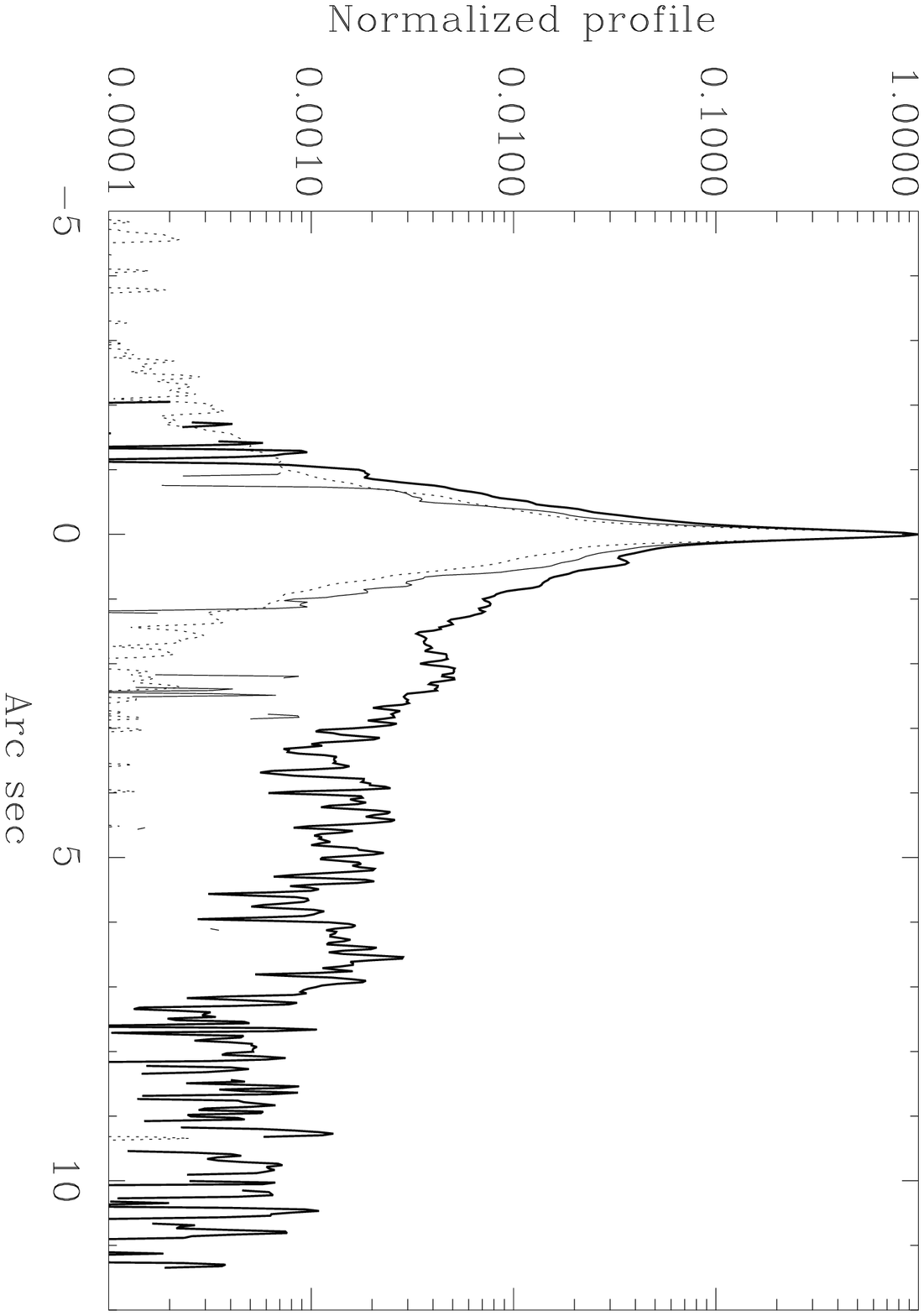}
\end{figure}

\begin{figure}
\plotone{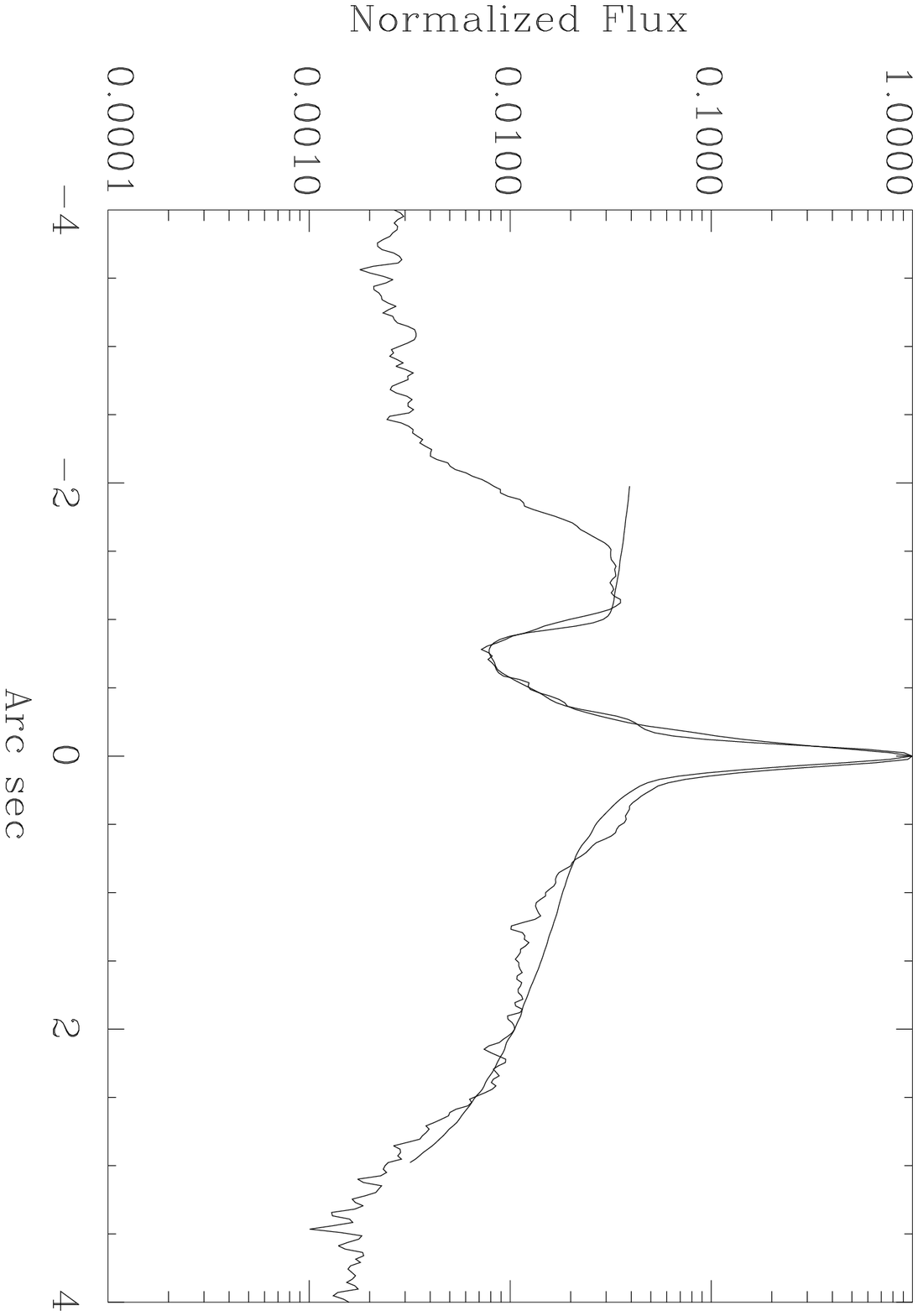}
\end{figure}

\begin{figure}
\plotone{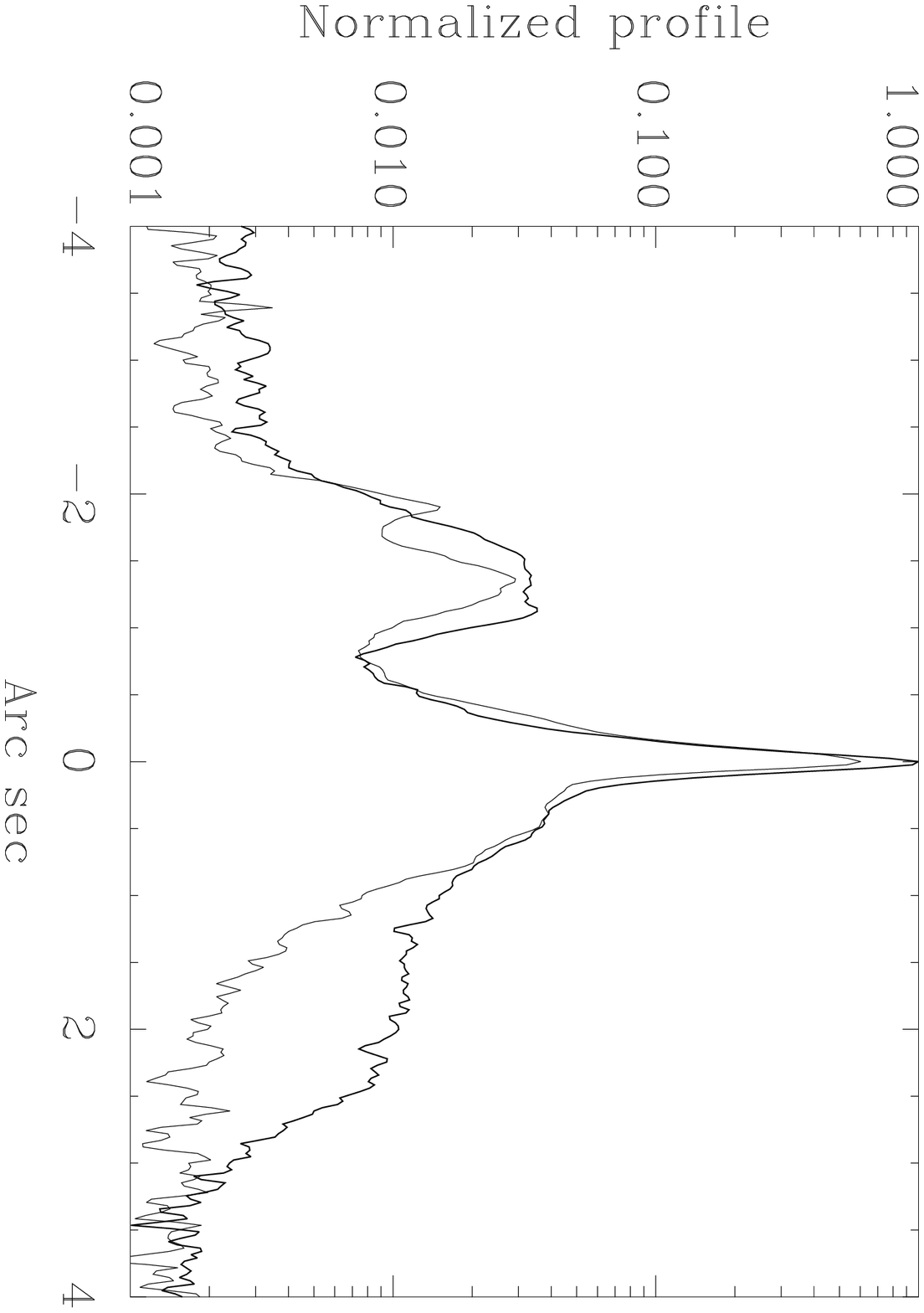}
\end{figure}

\begin{figure}
\plotone{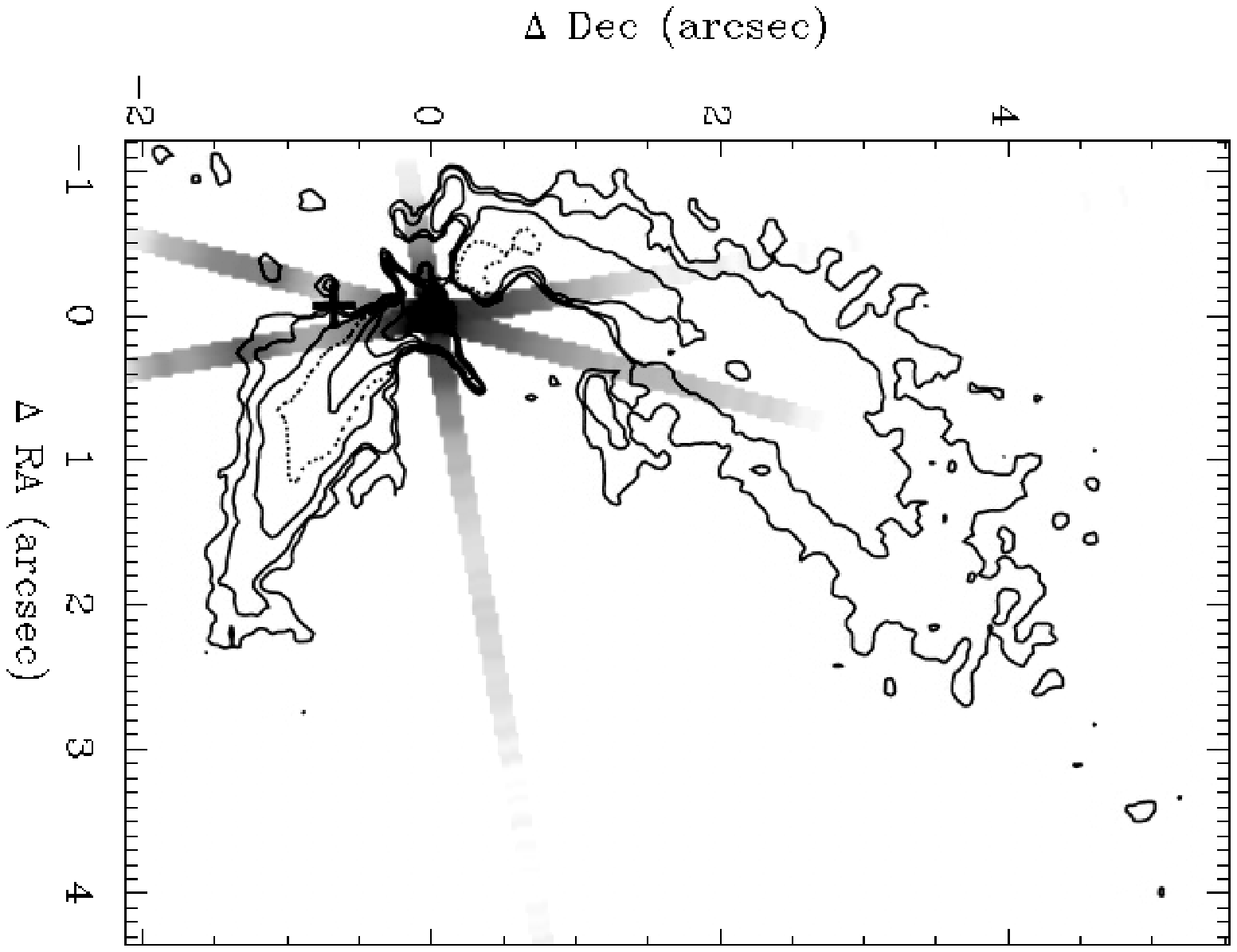}
\end{figure}

\end{document}